\documentclass[prx,twocolumn,superscriptaddress,amsmath,amssymb,floatfix]{revtex4-1}

\usepackage{bm}
\usepackage{graphicx}
\usepackage{color}
\usepackage{amsmath}
\usepackage{hyperref}
\usepackage{epstopdf}
\usepackage{upgreek}
\begin{document}

\title{Life and death of a cold BaRb$^+$ molecule inside an ultracold cloud of Rb atoms}

\author{Amir Mohammadi}
\affiliation{Institut f\"{u}r Quantenmaterie and Center for Integrated Quantum Science and
	Technology (IQ$^{\text{ST}}$), Universit\"{a}t Ulm, 89069 Ulm, Germany}
\author{Artjom Kr\"{u}kow}
\affiliation{Institut f\"{u}r Quantenmaterie and Center for Integrated Quantum Science and
	Technology (IQ$^{\text{ST}}$), Universit\"{a}t Ulm, 89069 Ulm, Germany}
\author{Amir Mahdian}
\affiliation{Institut f\"{u}r Quantenmaterie and Center for Integrated Quantum Science and
	Technology (IQ$^{\text{ST}}$), Universit\"{a}t Ulm, 89069 Ulm, Germany}
\author{Markus Dei\ss}
\affiliation{Institut f\"{u}r Quantenmaterie and Center for Integrated Quantum Science and
	Technology (IQ$^{\text{ST}}$), Universit\"{a}t Ulm, 89069 Ulm, Germany}
\author{Jes\'us P\'erez-R\'ios}
\affiliation{Fritz-Haber-Institut der Max-Planck-Gesellschaft, Faradayweg 4-6, 14195 Berlin, Germany}
\author{Humberto da Silva Jr.}
\affiliation{Universit\'{e} Paris-Saclay, CNRS, Laboratoire Aim\'{e} Cotton, 91405 Orsay Cedex, France}
\author{Maurice Raoult}
\affiliation{Universit\'{e} Paris-Saclay, CNRS, Laboratoire Aim\'{e} Cotton, 91405 Orsay Cedex, France}

\author{Olivier Dulieu}
\affiliation{Universit\'{e} Paris-Saclay, CNRS, Laboratoire Aim\'{e} Cotton, 91405 Orsay Cedex, France}
\author{Johannes Hecker Denschlag}
\affiliation{Institut f\"{u}r Quantenmaterie and Center for Integrated Quantum Science and
	Technology (IQ$^{\text{ST}}$), Universit\"{a}t Ulm, 89069 Ulm, Germany}
\date{\today}

\begin{abstract}
We study the evolution of a single BaRb$^+$ molecule while it continuously collides with ultracold Rb atoms. The initially weakly-bound molecule can undergo a sequence of elastic, inelastic, reactive, and radiative processes. We investigate these processes by developing methods for discriminating between
different ion species, electronic states, and kinetic ion energy ranges. By comparing the measurements to model calculations we obtain a consistent description of the typical trajectory of the ion through the manifold of available atomic and molecular states. As a further result, we determine rates for collisional and radiative relaxation as well as photodissociation, spin-flip collisions, and chemical reactions.
\end{abstract}

\maketitle

\section{Introduction}

In recent years, methods have been developed to produce ultracold molecules out of ultracold atoms, e.g. by photoassociation \cite{Jones2006,Hutson2006,Ulmanis2012}, sweeping over a Feshbach resonance \cite{Kohler2006,Chin2010}, radiative association in a two-body collision (e.g. \cite{Hall2011,Silva2015}), or three-body recombination \cite{Moerdijk1996, Fedichev1996, Burt1997, Stamper1998, Greene2017}. Typically, the resulting cold molecules are internally highly-excited and very reactive. Therefore, several questions arise. What are the reaction and relaxation paths that the particles take while they are exposed to light fields and collisions? What are the dynamics?

 Investigations on these topics can be conveniently performed in hybrid-atom-ion systems where trapped, cold molecular ions are immersed in a trapped gas of ultracold atoms \cite{Meyer2017, Zhang2017,Heazlewood2015,Mikosch2010,Gerlich2008a, Tomza2019, Haerter2014}. Ion traps can be very deep so that an ion is still trapped even if large amounts of energy are released in an inelastic or reactive process. Furthermore, it is possible to selectively detect ionic products on the single particle level. Control over the locations of the traps allows for deterministically starting or stopping collisional dynamics between atoms and ion. In addition, low temperatures in the mK regime and below enable a high level of control for the preparation of the initial quantum state of the reactants and of the collision parameters such as the collision energy.
A specific property of ion-neutral collisions is the long-range interaction between a charge and an induced dipole, which depends on the interatomic distance as $1/R^4$ \cite{Haerter2014, Tomza2019}. The combination of long-range interaction and low temperature corresponds to an interesting
regime where reactions and inelastic processes can already take place at comparatively large inter-particle distances (see, e.g., \cite{Jachymski2020}). This leads to large cross sections and promotes the formation of weakly-bound molecular states.

The young field of cold hybrid-atom-ion systems has shown tremendous progress studying inelastic collisions and reactions. This includes charge exchange between atoms and atomic ions \cite{Grier2009,Schmid2010,Ravi2012,Ratschbacher2012,Goodman2015,Haze2015,Joger2017}, and spin flips \cite{Sikorsky2018,Furst2018}. It was possible to observe collisionally induced vibrational or rotational relaxation of a deeply-bound molecular ion \cite{Rellergert2013, Hauser2015}, which is a collision at short internuclear distances. Furthermore, the formation of cold molecular ions from cold neutral and electrically charged atoms has been realized for several species (e.g., \cite{Hall2011,  Haerter2012, Hall2013, Sullivan2011}), and reactive behavior of molecular ions has been investigated \cite{Hall2012, Deiglmayr2012, Puri2017, Puri2019}.

Here we take a different approach, focussing less on a single, particular physical or chemical process.
Instead we study the progression and interplay of the elastic, inelastic and reactive processes which take place.
Concretely, we investigate, both experimentally and theoretically,
the evolution of a cold, weakly-bound BaRb$^+$ molecular ion as it continuously collides with ultracold Rb atoms. These collisions can be elastic, inelastic, or reactive. Our investigation includes the deterministic birth of the molecular ion inside the atom cloud, its typical life undergoing changes in the electronic and vibrational states, and its death as it reacts away. We find that the evolution of the BaRb$^+$ ion directly after its formation is mainly dominated by vibrational relaxation collisions with Rb atoms at large internuclear distance. With increasing binding energy, radiative processes become progressively important until they are dominant. We observe Ba$^+$, Rb$_2^+$ and Rb$^+$ ions as reaction products, resulting from a range of photo- or collisionally-induced processes which are discussed in detail. Interestingly, in the experiments of Ref.$\:$\cite{Hall2013} where the formation of BaRb$^+$ molecules from cold Ba$^+$ ions and Rb atoms was studied, also Rb$^+$ and Rb$_2^+$ as final products were detected. How these products came about, however, remained unclear. The results of our work, presented here, may be a key to also explain these findings.

This article is organized as follows. In Secs.$\:$\ref{sec:setup} to \ref{sec:conclusion}, we study the elastic, inelastic and reactive processes of the BaRb$^+$ ion for different phases of its evolution. A detailed discussion of experimental parameters and detection methods is provided in Secs.$\:$\ref{sec:app_CloudDensity} to \ref{sec:app_Detection_ion} of the Appendix. Finally, in Secs.$\:$\ref{sec:app_ColCrossSec} and \ref{sec:App_MC} of the Appendix we give additional information on the theoretical models, calculations, and Monte-Carlo (MC) simulations.

\begin{figure}
\includegraphics[width=8.5cm] {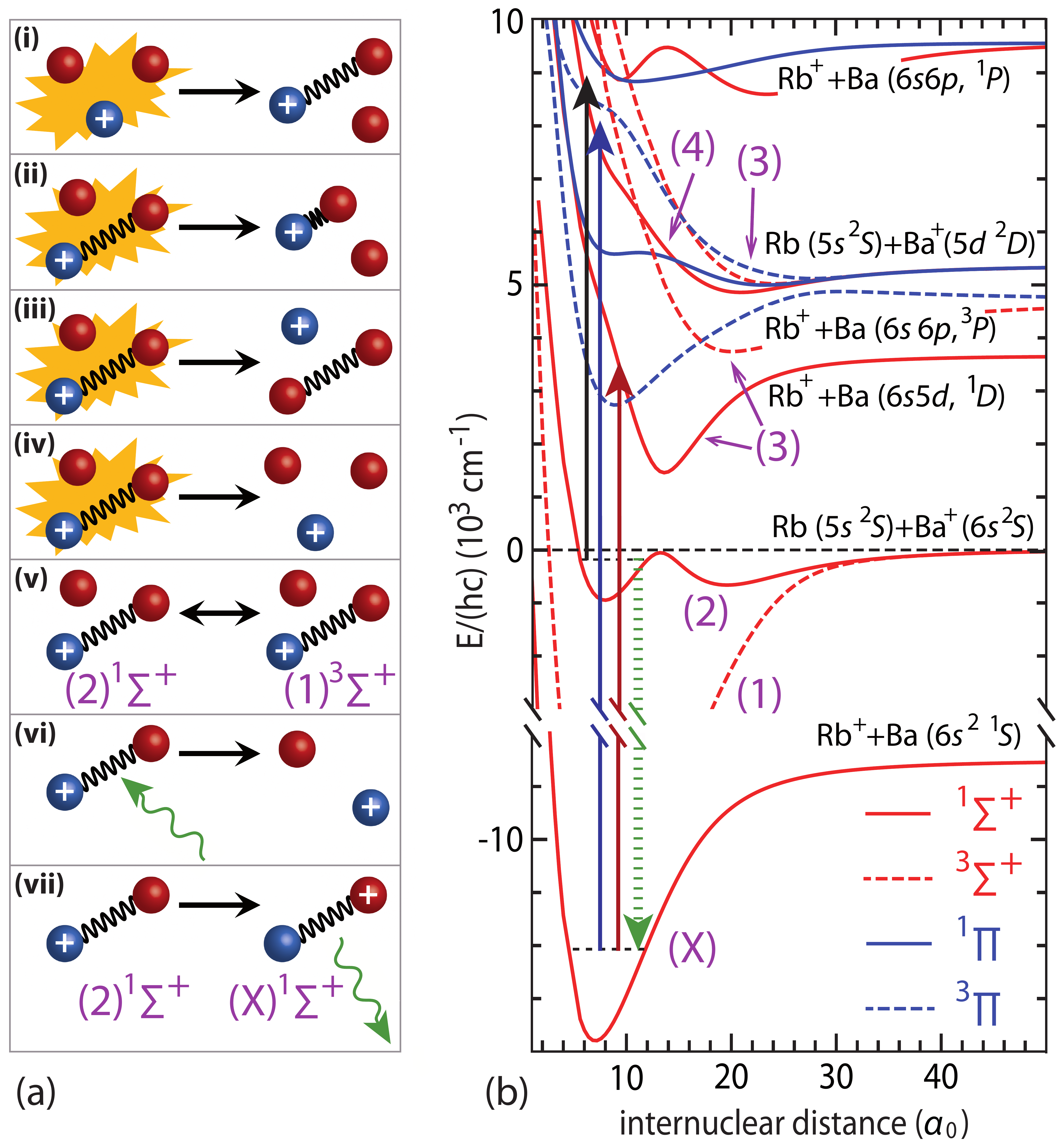}
\caption{(a) Illustration of various inelastic and reactive processes. (i) Formation of a BaRb$^+$ molecule via three-body recombination, (ii) collisional relaxation of a BaRb$^+$, (iii) substitution reaction, (iv) collisional dissociation, (v) collisional spin exchange, (vi) photodissociation, and (vii) radiative relaxation. (b) PECs for BaRb$^+$, taken from \cite{Hall2013}. The entrance channel $\text{Rb}(5s\,^2S)+\text{Ba}^+(6s\,^2S)$ marks zero energy. Solid black, blue, and red arrows show possible photodissociation transitions for 1064$\:$nm, 493$\:$nm, and 650$\:$nm light, respectively. The dashed green arrow indicates radiative relaxation to the electronic ground state.}
\label{fig1}
\end{figure}

\section{Experimental setup and production of molecular ion}
\label{sec:setup}
Our experiments are carried out in a hybrid atom-ion apparatus. The basic setup is described in detail in \cite{Schmid2012}. For the investigations presented here, we produce a single BaRb$^+$ molecule which is trapped in a linear Paul trap with trap frequencies of $2\pi\times(80,\:30)\:$kHz in radial and axial direction, respectively. The BaRb$^+$ ion is in contact with a cloud of $6\times10^6$ ultracold $^{87}$Rb atoms with a temperature of $T=750\:$nK. The atoms are prepared in the electronic ground state $5S_{1/2}$ and are spin-polarized, having a total angular momentum $F=1$ and $m_F=-1$. The atomic cloud is held in a far off-resonant crossed optical dipole trap (ODT) at 1064$\:$nm with a trap depth of approximately 20$\:\upmu \text{K}\times\text{k}_\text{B}$, where $k_\text{B}$ is the Boltzmann constant. The density distribution of the cigar-shaped cloud can be described by a Gaussian with root mean square widths of 9 and 60$\:\upmu$m in radial and axial direction, respectively (see Appendix \ref{sec:app_CloudDensity}).

Initially, the cold BaRb$^+$ molecule is produced via three-body recombination $\text{Ba}^++\text{Rb}+\text{Rb}\rightarrow \text{BaRb}^++\text{Rb}$, typically at large internuclear distances \cite{Kruekow2016a,Kruekow2016b}, see (i) in Fig.$\:$\ref{fig1}(a). For this, we prepare in the Paul trap a single, laser-cooled $^{138}$Ba$^+$ ion in the electronic ground state 6$S_{1/2}$, and a dense Rb atom cloud in the ODT. At that time the two traps are separated by about 100$\:\upmu$m. Right before we start our experiments with the single Ba$^+$ ion we remove unwanted Rb$^+$ and Rb$_2^+$ ions, which can form spontaneously in our trapped atom cloud \cite{NoteSpontIon}, with a mass-filter scheme, see Appendix \ref{sec:Massfilter}.
After this purification step the 493$\:$nm and 650$\:$nm laser-cooling beams for the Ba$^+$ ion are switched off and the Ba$^+$ ion is moved into the atom cloud center. This is done within 100$\:\upmu$s by abruptly changing the voltage on one of the endcap electrodes of the Paul trap by $1.5\:$V.

Once the Ba$^+$ ion is in the atom cloud the $\text{Ba}^++\text{Rb}+\text{Rb}\rightarrow \text{BaRb}^++\text{Rb}$ three-body recombination leads to the formation of BaRb$^+$ molecules with a rate $\Gamma_\text{tbr}=k_3 n(t)^{2}$, where $k_3 = 1.04(4) \times 10^{-24}\:\text{cm}^{6}\text{s}^{-1}$ is the three-body rate constant \cite{Kruekow2016a}, and $n(t)$ is the density of the atom cloud at a given time $t$ at the ion trap center. For the central atomic density of $8.1\times 10^{13}$cm$^{-3}$ we obtain $\Gamma_\text{tbr}\approx 6.8\times 10^3\:\text{s}^{-1}$. Three-body recombination is by orders of magnitude the leading reaction process of the Ba$^+$ ion, and BaRb$^+$ is the main product \cite{Kruekow2016b, Kruekow2016a}. Initially, the BaRb$^+$ molecule is weakly-bound below the atomic $\text{Rb}(5s\,^2S)+\text{Ba}^+(6s\,^2S)$ asymptote [see Fig.$\:$\ref{fig1}(b)]. Its binding energy is expected to be $\sim2 \text{mK}\times k_\text{B}$ corresponding to the typical atom-ion collision energy in our Paul trap \cite{Kruekow2016a,Kruekow2016b}. Furthermore, according to simple statistical arguments, we expect the BaRb$^+$ molecular ion to be produced in the singlet state $(2)^1\Sigma^+$ and triplet state $(1)^3\Sigma^+$ with a probability of 25\% and 75\%, respectively. For both of them the initial binding energy of $\sim2 \text{mK}\times k_\text{B}$ corresponds to a vibrational state $v=-5$ (see also Fig.$\:$\ref{fig:Ebinds} of Appendix \ref{sec:QTC}). The negative vibrational quantum number $v$ indicates that it is counted downwards from the atomic asymptote, starting with $v = -1$ for the most weakly-bound vibrational state. When $v$ has a positive value, it is counted upwards from the most deeply-bound vibrational state $v = 0$.

\section{Experimental Investigation of the evolution of the molecular ion}
\begin{figure}[t]
	\includegraphics[width=\columnwidth] {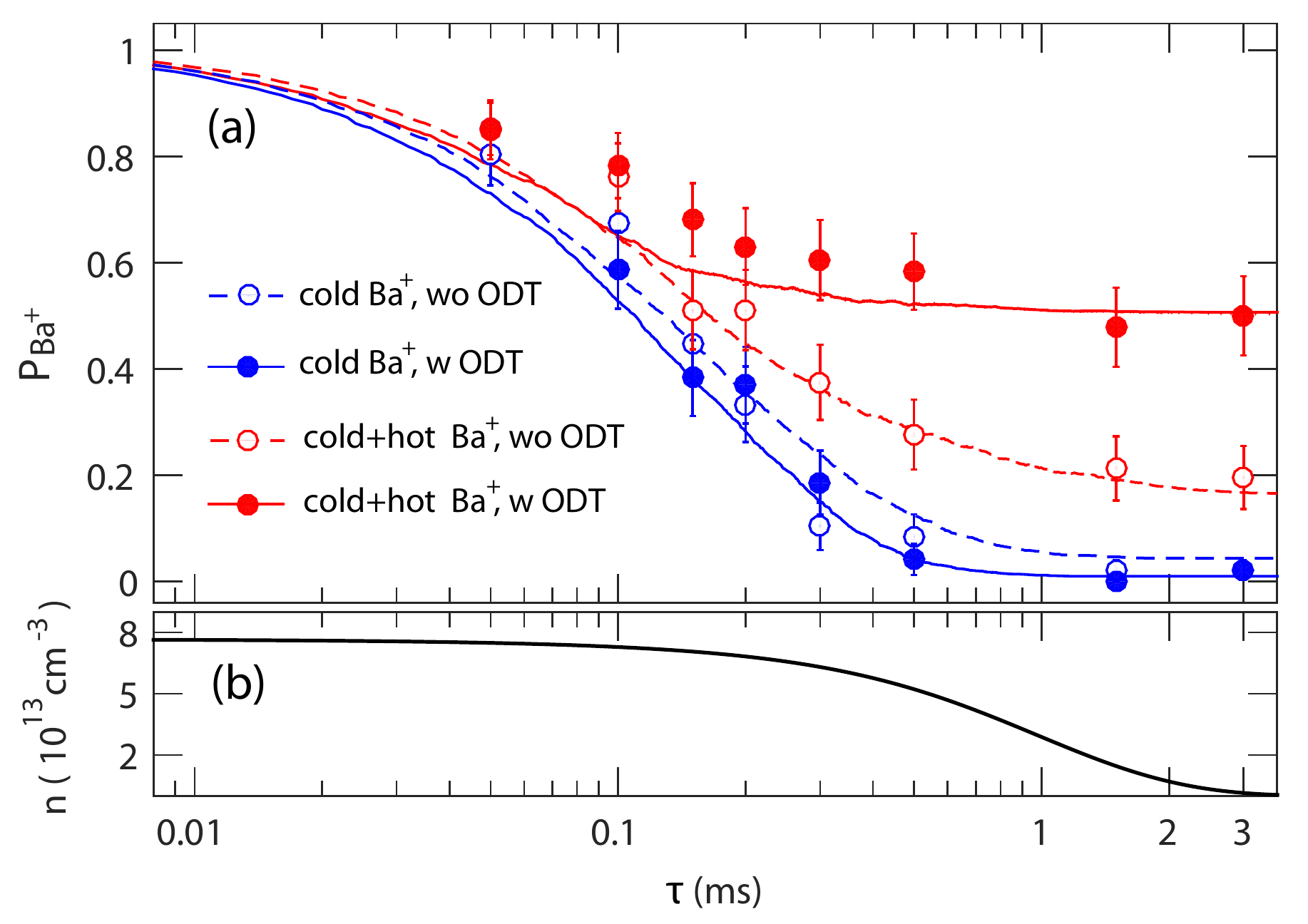}
	\caption{(a) Probability $P_{\text{Ba}^+}$ of detecting a Ba$^+$ ion as a function of time $\tau$ after immersion into a Rb atomic cloud. Circles are measured data. Each data point is the mean value of 50 repetitions of the experiment and the error bars represent the 1$\sigma$ statistical uncertainty. Curves are the results of MC simulations (see Appendix \ref{sec:App_MC}). (b) Time evolution of the atomic density $n$ at the ion trap center after the ODT beams have been switched off at $\tau=-250\:\upmu$s.
	}
	\label{fig2}
\end{figure}

As will become clear later, we can learn a lot about the evolution of the BaRb$^+$ molecule by monitoring the presence of the Ba$^+$ ion and its state in the trap. Figure \ref{fig2}(a) shows data for the measured probability $P_{\text{Ba}^+}$ for detecting a Ba$^+$ ion as a function of time for four different experiments. After immersing the cold Ba$^+$ ion into the cloud for a variable time $\tau$ we quickly (within 20$\:\upmu$s) pull out the remaining ion and take two fluorescence images (see Appendix \ref{sec:Fluores_img} for details). For the first image the imaging parameters are chosen such, that only a cold Ba$^+$ ion with a temperature of about $100\,\text{mK}$ or below can be detected. The filled blue circles in Fig.$\:$\ref{fig2}(a) show
these measurements for various immersion times $\tau$.
We essentially observe here the three-body recombination of Ba$^+$ towards BaRb$^+$. Next, we take a second fluorescence image which is preceded by a long laser cooling stage (for details see Appendix \ref{sec:Fluores_img}). This retrieves almost 60\% of the Ba$^+$ ions that had reacted away [filled red circles in Fig.$\:$\ref{fig2}(a)].
We can explain this retrieval by the following scenario. There is a sizable probability for a freshly formed BaRb$^+$ molecular ion to break up via photodissociation.
The break up produces a hot Ba$^+$ ion which is subsequently cooled down to below $\approx 100\:$mK by the long laser cooling stage so that it can be detected by fluorescence imaging (see Appendix \ref{sec:Fluores_img}). Photodissociation can occur, e.g., by the ODT laser at 1064 nm.
Figure$\:$\ref{fig1}(b) shows indeed that 1064$\:$nm photons can excite weakly-bound BaRb$^+$ ions below the $\text{Rb}(5s\,^2S)+\text{Ba}^+(6s\,^2S)$ asymptote to repulsive potential energy curves (PECs).  The most relevant transitions to produce a hot Ba$^+$ ion are
$(2)^1\Sigma^+ \rightarrow (4)^1\Sigma^+$ and $(1)^3\Sigma^+ \rightarrow (3)^3\Pi$.
 After the excitation, the Ba$^+$ ion and the Rb atom are accelerated away from each other, following the repulsive molecular potential. The Ba$^+$ ion will obtain a high kinetic energy of up to 0.2$\:$eV. As a consequence, it will afterwards orbit most of the time outside the atom cloud, having a small probability for collisions with Rb atoms. Therefore, sympathetic cooling and three-body recombination are strongly suppressed and the hot Ba$^+$ ion remains hot until it is cooled down during the long laser cooling stage.
We have direct evidence for this photodissociation process, since we detect
 a fraction of the  Ba$^+$ ions in the electronically excited $5D_{5/2}$ state which corresponds to one of the asymptotic states of the $(4)^1\Sigma^+$
and $(3)^3\Pi$ potentials [see Fig.\ref{fig1}(b)].  Concretely, we find that about half of the retrieved hot Ba$^+$ ions populate the metastable $5D_{5/2}$ state with its natural lifetime of $\sim 30\:$s.
As the $\text{Rb}(5s)+\text{Ba}^+(5d \, D_{3/2, 5/2}$) asymptotes are located  more than $5000\:\text{cm}^{-1}$ above the initially formed BaRb$^+$ molecular states, they only can be reached by photodissocation. We discriminate the population of the Ba$^+$ $5D_{5/2}$ state from the population in the other Ba$^+$ states by using the fact, that a Ba$^+$ ion in state
$5D_{5/2}$ can only be laser-cooled and detected after pumping it out of this metastable level with a $614\:\text{nm}$ laser. Thus, when we switch off the $614\:\text{nm}$ laser we lose the signal from the metastable $5D_{5/2}$ Ba$^+$ ion.

To double check whether it is really the 1064$\:$nm ODT laser which is responsible for photodissociation we carry out a second set of measurements, where the ODT is turned off 250$\:\upmu$s before the Ba$^+$ ion is immersed into the cold atom cloud. As a consequence the atomic cloud is now free falling and ballistically expanding. The calculated time evolution of the atomic density at the center of the ion trap is shown in Fig.$\:$\ref{fig2}(b) (see also Appendix \ref{sec:app_CloudDensity}). When we detect cold Ba$^+$ ions via fluorescence imaging [hollow blue circles in Fig.$\:$\ref{fig2}(a)] there is essentially no change in signal as compared to the case with the ODT being on. This is expected since the atomic density is nearly constant on the time scale of the three-body recombination.  However, the signal solely for the hot Ba$^+$ ion, which is obtained by subtracting the signal for cold Ba$^+$ from the signal for both cold+hot Ba$^+$, is significantly smaller compared to when the ODT laser is on.  Thus, this indeed shows that $1064\,\text{nm}$ light photodissociates BaRb$^+$ molecules into hot Ba$^+$ ions and Rb atoms.  Nevertheless, the signal for the hot Ba$^+$ ion is still on the order of 10\% for sufficiently large times $\tau$. Therefore, also light with a different wavelength than 1064$\:$nm must contribute to the production of hot Ba$^+$ ions. As we will show in Sec.$\:$\ref{sec:Photo_ground} the remaining signal for hot Ba$^+$ can be explained due to photodissociation of ground state $(X)^1\Sigma^+$ molecules via the laser cooling light at 493$\:$nm.

We note that photodissociation by $1064\:\text{nm}$ light can also produce a hot Rb$^+$ ion, instead of a hot Ba$^+$ ion. This occurs in the transition
$(1)^3\Sigma^+ \rightarrow (3)^3\Sigma^+$. So far, we have not experimentally studied this process in detail.

\section{Insights from calculations}
\label{sec:insights}

We now combine the information from our experimental data with insights from theoretical calculations.  This sets strong constraints on possible scenarios for the evolution of the BaRb$^+$ molecule and essentially fixes all free parameters of our theoretical model. Our analysis mainly involves electronic and vibrational states, while rotational and hyperfine degrees of freedom are not taken into account to a large part.

For example, we have carried out calculations for photodissociation cross sections which are
based on computed PECs and transition dipole moments for highly-excited electronic states \cite{Hall2013,Silva2015} (for details see Appendix \ref{sec:PhotoDisso}).
 In the calculations we find that the photodissociation cross section for the $v = -5$ BaRb$^+$ molecule with 1064$\:$nm light  is about two orders  of magnitude too small to explain the hot Ba$^+$ signal. However, the calculations also show that the photodissociation cross section increases approximately as $\propto E_b^{0.75}$ for both singlet $(2)^1\Sigma^+$ and triplet $(1)^3\Sigma^+$ BaRb$^+$ molecules, where $E_b$ is the binding energy (see Fig.$\:\:$\ref{fig:CrossSectionVsBinEnergy} of Appendix \ref{sec:PhotoDisso}). Apparently, shortly after production, while it is still immersed in the Rb cloud, the weakly-bound BaRb$^+$ molecule must vibrationally relax by a number of vibrational levels before it is photodissociated. A theoretical treatment shows that the vibrational relaxation is due to inelastic atom-molecule collisions, for which we have derived cross sections in Appendix \ref{sec:QTC} via quasi-classical trajectory (QCT) calculations. Furthermore, our calculations predict that while photodissociation of singlet $(2)^1\Sigma^+$ molecules via the 1064$\:$nm laser indeed dominantly produces hot Ba$^+$ ions, photodissociation of triplet $(1)^3\Sigma^+$ molecules mainly leads to hot Rb$^+$ ions. Thus, in order to explain the measured substantial percentage of hot Ba$^+$ ions, there has to be a mechanism which converts triplet molecules into singlet molecules. This spin-flip mechanism is provided by inelastic atom-molecule collisions, for which the cross section is estimated to be a fraction of the Langevin cross section (see Appendix \ref{ssec:spin-flip}). Finally, our theoretical treatment reveals that radiative relaxation of the $(2)^1\Sigma^+$ molecules towards the electronic ground state $(X)^1\Sigma^+$ due to spontaneous emission [as illustrated by the green downward arrow in Fig.$\:$\ref{fig1}(b)] needs to be taken into account. According to our calculations we obtain a broad population distribution of final vibrational levels in the ground state, ranging from about $v=10$ to above $v=200$, with a peak at $v=55$, see Fig.$\:$\ref{fig:vib_xbarb} in Appendix \ref{sec:PhotoDisso}.
 For a relaxation towards the $v = 55$ level a photon at a wavelength of about 850$\:$nm is emitted. The relaxation rate is predicted to  scale as $\propto E_b^{0.75}$, which is the same power law as for photodissociation, see Fig.$\:$\ref{fig:LifetimeVsBinEnergy} in Appendix \ref{sec:PhotoDisso}. Hence, there is a constant competition between photodissociation and radiative relaxation for the singlet state $(2)^1\Sigma^+$. Once in the ground state the molecule is immune to photodissociation by 1064$\:$nm light, because the photon energy is not sufficient. Photodissociation via laser cooling light, however, is possible.

 We note that triplet $(1)^3\Sigma^+$ molecules cannot radiatively relax to $(X)^1\Sigma^+$ according to the selection rules for electric dipole transitions. Therefore, in the absence of any collisional or light-induced processes, these molecules remain within the triplet state $(1)^3\Sigma^+$.

Besides the already mentioned inelastic and reactive processes also collisional dissociation, substitution reactions and elastic collisions play a role for the evolution of the BaRb$^+$ molecule.
  In order to theoretically model the evolution of the BaRb$^+$ molecular ion in the atom cloud
we carry out MC simulations  and compare them to the measured data.
  In the simulations we take into account the most relevant processes shown in Fig.$\:$\ref{fig1}(a), as well as additional ones.
  Details on the simulations, the various processes, and their respective cross sections can be found in the
  Appendices \ref{sec:app_ColCrossSec} and \ref{sec:App_MC}.
  The simulations produce the  lines in Fig.$\:$\ref{fig2}(a), showing reasonable agreement with the experimental data. Additional results of the calculations can be found in  Fig.$\:$\ref{fig:all_MC_results} in Appendix \ref{sec:fullevolution}.

\section{Evolution of the molecular ion}

In the following, we discuss the results of our analysis. Our theoretical investigations  show that the evolution of the BaRb$^+$ molecule both for the singlet state $(2)^1\Sigma^+$ and the triplet state $(1)^3\Sigma^+$ will at first be dominated by vibrational relaxation collisions, which occur approximately with the Langevin rate $\Gamma_L= 164\:\text{ms}^{-1}$ for the peak atomic density of $8.1\times 10^{13}$cm$^{-3}$ in our cloud (see Appendix \ref{sec:QTC}). Typically, these collisions lead to vibrational relaxation in steps of one or two vibrational quanta, with an average of 1.4 vibrational quanta per Langevin collision (see Appendix \ref{subs:Evolution}). Vibrational relaxation heats up the ion since binding energy is released in form of kinetic energy. This is counteracted by sympathetic cooling due to elastic collisions with Rb atoms, which occur at an average rate of about one elastic collision (with sizable momentum exchange) per vibrational relaxation step (see Appendix \ref{subs:Evolution}). As a consequence the typical temperature of the BaRb$^+$ ion is below $15\:\text{mK}$ during the initial, collision-dominated phase of the evolution.  In general, when the collision energy is larger than the binding energy of the BaRb$^+$ molecule, the molecule can also dissociate into a (cold) Ba$^+$ ion and a Rb atom. For the initial vibrational level $v=-5$ with its binding energy of $2\:\text{mK}\times k_\text{B}$ this process occurs, however, only with a comparatively small rate of about $\Gamma_L/7$ (see Appendix \ref{sec:QTC}), and is negligible for deeper vibrational levels. Furthermore, our calculations reveal that for weakly-bound BaRb$^+$ ions in the states $(2)^1\Sigma^+$ and $(1)^3\Sigma^+$ the rate for the substitution reaction $\text{BaRb}^++\text{Rb}\rightarrow \text{Rb}_2+\text{Ba}^+$ is negligible. This is a consequence of the fact that the interaction between the Rb atoms is much more short range than between a Rb atom and the Ba$^+$ ion, see also Appendix \ref{sec:QTC}. Concerning the spin-flip collisions we obtain good agreement with the experimental data when using a spin-flip rate of $\Gamma_L/42$ for flips from triplet to singlet (see Appendix \ref{ssec:spin-flip}).

Our calculations predict that for the experiments with ODT, which is operated at an intensity of $18\:\text{kW}\,\text{cm}^{-2}$, the $(2)^1\Sigma^+$ molecules vibrationally relax typically to a level $v = -12$ ($E_b\approx50\:\text{mK}\times k_\text{B}$) before either photodissociation or radiative relaxation to the ground state. By contrast, $(1)^3\Sigma^+$ molecules, which cannot radiatively relax to the ground state, typically reach a deeper vibrational level of $v = -21$ ($E_b\approx460\:\text{mK}\times k_\text{B}$).  The photodissociation rates are given by $\Gamma_\text{PD}=\sigma_\text{PD} I/(h\nu)$, where $I$, $h$, and $\nu$ are the laser intensity, the Planck constant, and the laser frequency, respectively, and $\sigma_\text{PD}$ is the photodissociation cross section. Calculations for $\sigma_\text{PD}$ are presented in Appendix \ref{sec:PhotoDisso}. For the experiments without ODT (and thus without corresponding photodissociation channel) $(2)^1\Sigma^+$ molecules are expected to relax typically to $v = -18$ ($E_b\approx230\:\text{mK}\times k_\text{B}$) before radiative relaxation to the ground state takes place.

\begin{figure}
	\includegraphics[width=\columnwidth] {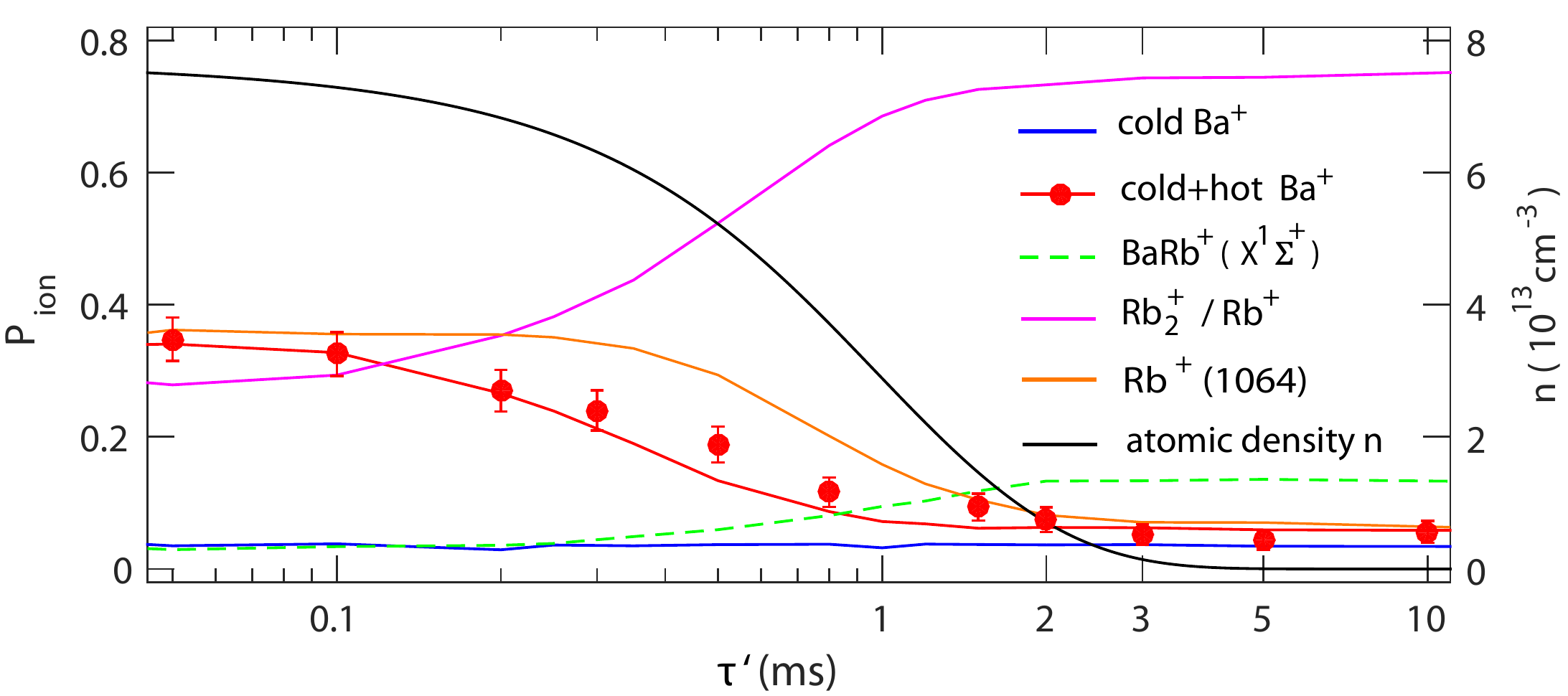}
	\caption{Observation of radiative relaxation. The data points give the measured probability $P_{\text{Ba}^+}$ of detecting a Ba$^+$ ion after an extensive laser cooling stage as a function of the time $\tau'$ after which a 300$\:$ms long 1064$\:$nm light pulse is applied. For formation of a BaRb$^+$ molecule, a cold Ba$^+$ ion is immersed into the atom cloud at $\tau' = 0$. The solid black line represents the time evolution of the atomic density $n$ at the ion trap center (see Appendix \ref{sec:app_CloudDensity}). All other lines are results of MC calculations for the probabilities of finding the species as denoted in the legend.
	}
	\label{fig3}
\end{figure}

\section{Radiative relaxation}
\label{sec:relaxation}

Since radiative relaxation to the ground state is predicted to be a central process in the evolution of the BaRb$^+$ ion, we now test for it experimentally. The idea is to measure for how long BaRb$^+$ molecules remain in the excited states $(2)^1\Sigma^+$ or $(1)^3\Sigma^+$ before they radiatively relax to the ground state $(X)^1\Sigma^+$. We probe the presence of a BaRb$^+$ molecule in the states $(2)^1\Sigma^+$ or $(1)^3\Sigma^+$ by photodissociating it into a Ba$^+$ ion and a Rb atom with the 1064$\:$nm laser, and then detecting the hot Ba$^+$ ion.  BaRb$^+$ molecules in the ground state $(X)^1\Sigma^+$ cannot be photodissociated by the 1064$\:$nm laser because the photon energy is not sufficient.
We start this experiment by moving a single and cold Ba$^+$ ion into the atom cloud 250$\:\upmu$s after the 1064$\:$nm ODT laser has been switched off. As before, a BaRb$^+$ molecule will form on a time scale of $\Gamma_\text{tbr}^{-1}=0.15\:\text{ms}$. After the immersion of the Ba$^+$ ion, we wait for a time $\tau'$ before we switch on again the 1064$\:$nm laser
\cite{NoteVertLaser}
 to photodissociate the molecule.
Before applying the detection scheme for the released hot Ba$^+$ ion, i.e. long laser cooling and subsequent fluorescence imaging, we remove any remaining BaRb$^+$ molecule by  mass-filtering (see Appendix \ref{sec:Massfilter}).  The removal is done, because a remaining BaRb$^+$ molecule can give rise to a spurious hot Ba$^+$ signal  as the laser cooling step can also photodissociate a BaRb$^+$ molecule into a Ba$^+$ ion and a Rb atom. This is discussed in detail later in Sec.$\:$\ref{sec:Photo_ground}.

The red data points in Fig.$\:$\ref{fig3} show the probability $P_{\text{Ba}^+}$ to detect a Ba$^+$ ion (hot or cold) at the end of the given experimental sequence for various times $\tau'$. As expected, the Ba$^+$ signal decreases as $\tau'$ increases because the BaRb$^+$ molecule has more time to relax to the $(X)^1\Sigma^+$ state.  The decrease to about $1/3$ of the initial value takes place within about $\tau'=0.5\:$ms, which represents an approximate time scale for the lifetime of the $(2)^1\Sigma^+$ and $(1)^3\Sigma^+$ BaRb$^+$ molecule, respectively, in the cloud of Rb atoms.
For times longer than 2$\:$ms  an almost constant value of $P_{\text{Ba}^+}\approx 6\%$
is observed.  This remaining population is composed of the following contributions: 4\%  are cold Ba$^+$ ions (blue solid line) that have not reacted at all
 \cite{NoteAtomOffset} or that have been released again as a result of collisional dissociation.
 $2\%$  arise probably from BaRb$^+$ molecules that are stuck in the triplet state $(1)^3\Sigma^+$ after the collisional phase when all neutral atoms have left for $\tau' > 2\:\text{ms}$, and are photodissociated from there by the 1064$\:$nm light.
   The green dashed curve gives the probability for ending up with a BaRb$^+$ molecule in the electronic ground state. This probability nearly reaches $P_{\text{BaRb}^+}= 20\%$. In principle, this fraction would be about four times as large, if the substitution reaction $\text{BaRb}^+(X) +\text{Rb} \rightarrow \text{Ba}+\text{Rb}_2^+$, which depletes electronic ground state BaRb$^+$(X) molecules, were absent
   \cite{NoteOtherReaction}.
The corresponding reaction rate is expected to be on the order of the Langevin rate (see Appendix  \ref{sec:QTC}).
    We note that a Rb$_2^+$ molecular ion can also decay in the collision
    $\text{Rb}_2^++\text{Rb} \rightarrow \text{Rb}^++\text{Rb}_2$,  if it is not too deeply bound \cite{NoteRb}.
    In our simulations, however, we do not further pursue this process, and therefore give here the joint probability for finding a Rb$_2^+$ ion or its  Rb$^+$  decay product (magenta line in Fig.$\:$\ref{fig3}). The orange curve, in contrast, gives the probability for Rb$^+$ ions which are produced via photodissociation by the $1064\,\text{nm}$ laser.

\section{Product ion species}

Our discussion so far already indicates that we expect to find a range of ionic products in our experiments, each  with a respective abundance.
We test this prediction by performing mass spectrometry in the Paul trap after the reactions. As described in more detail in Appendix \ref{sec:Massfilter}, this is done as follows. When probing for a given ionic product, we use the mass-filter to remove the ion from the Paul trap if it has the corresponding mass.  Afterwards, we check whether the Paul trap is now empty (see Appendix \ref{sec:app_Detection_ion}), knowing that before the mass-filtering a single ion of some species was present.
For the experiments without ODT we observe the following abundances after an interaction time $\tau = 10\:\text{ms}$: Ba$^+$: $4\pm2$\%, BaRb$^+$: $29\pm5$\%, Rb$_2^+$: $22\pm5$\%, Rb$^+$: $45\pm6$\%. Here, Rb$^+$ ions are probably created via the aforementioned reaction $\text{Rb}_2^++\text{Rb} \rightarrow \text{Rb}^++\text{Rb}_2$ as the direct process  $\text{BaRb}^++\text{Rb} \rightarrow \text{Rb}^++\text{BaRb}$ is slow and even energetically closed for  most vibrational states in the ground state $(X) ^1\Sigma^+$ (see Appendix \ref{sec:QTC}).  We expect the detected Ba$^+$ ions to be cold because there is no photodissociation light present in the given measurement scheme.  Our MC simulations are in good agreement with these abundances. From  Fig.\ref{fig:all_MC_results} we can
read off the following values for $\tau>3\,\text{ms}$: cold Ba$^+$: $\sim 4\%$, BaRb$^+$: $\sim 26\%$, Rb$_2^+$/Rb$^+$: $\sim70\%$,
  According to our simulations about $8\%$ of the initial Ba$^+$ ions end up as a BaRb$^+$ ion in the triplet state $(1)^3\Sigma^+$ while there are no molecules remaining in the singlet state $(2)^1\Sigma^+$ at $\tau = 10\:\text{ms}$. Therefore, the measured BaRb$^+$ fraction of 29\%  mainly consists of $(X)^1\Sigma^+$ electronic ground state molecules.

\begin{figure}[t]
	\includegraphics[width=8.5cm] {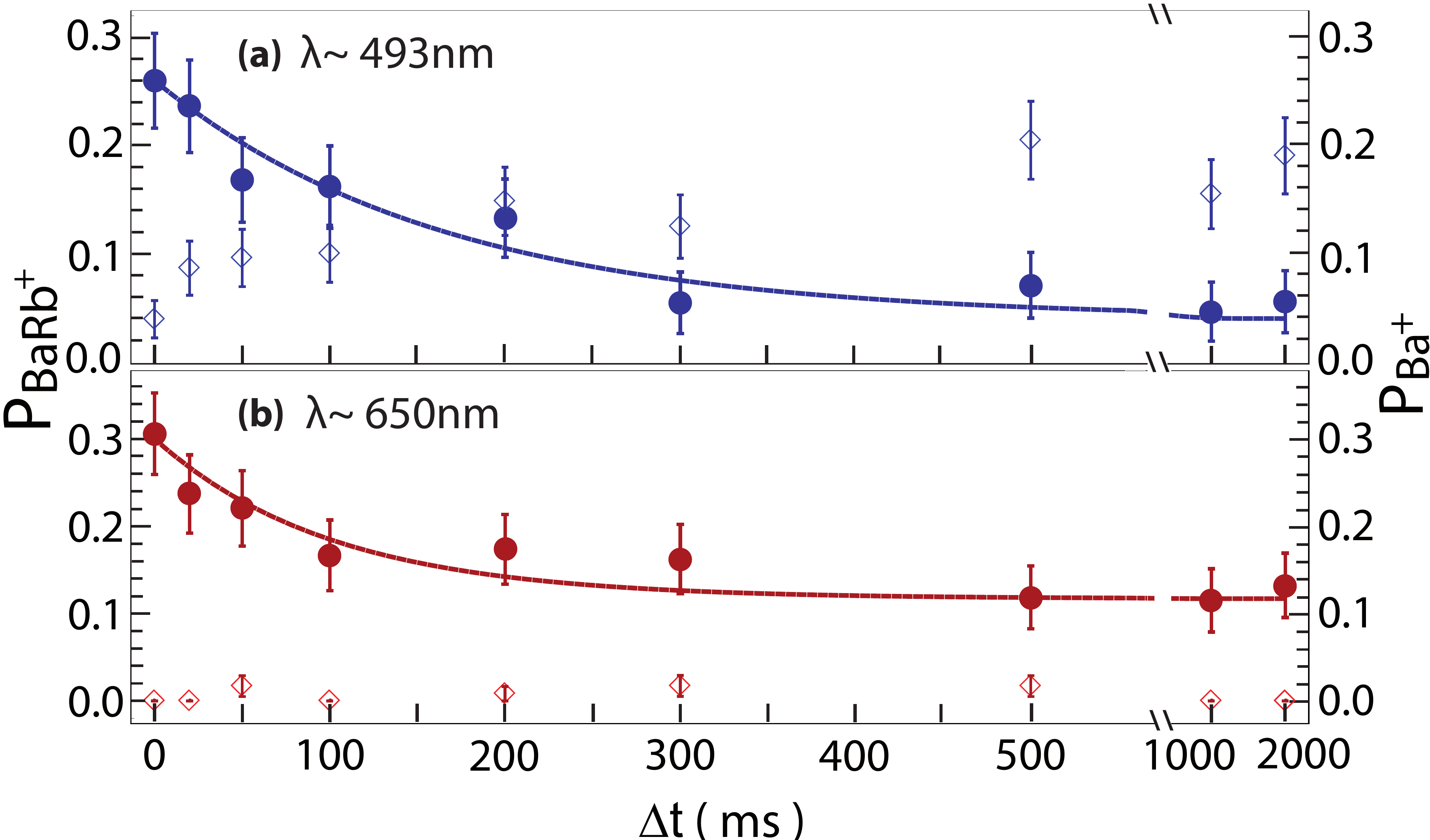}
	\caption{Photodissociation of ground state BaRb$^+$ molecules by 493$\:$nm (a) and 650$\:$nm (b) laser light as a function of exposure time $\Delta t$. Filled circles show the probability of detecting a BaRb$^+$ molecule. Hollow diamonds give the probability of detecting a Ba$^+$ ion. The solid lines are fits of an exponential decay plus offset.}
	\label{fig4}
\end{figure}

\section{Photodissociation of electronic ground state molecules}
\label{sec:Photo_ground}
Finally, we investigate photodissociation of the $(X)^1\Sigma^+$ state molecules. Once a BaRb$^+$ molecule has relaxed towards $(X)^1\Sigma^+$
 it is stable with respect to $1064\,\text{nm}$ light, however, photons from the cooling lasers for Ba$^+$ at the wavelengths of 650$\:$nm or 493$\:$nm can still photodissociate it, see blue and red arrows in Fig.$\:$\ref{fig1}(b) \cite{Note6}.  Figure$\:$\ref{fig4} shows photodissociation as a function of exposure time $\Delta t$ for light at 493$\:$nm (a) and at 650$\:$nm (b), respectively. The filled circles represent the fraction of experimental runs where we detect a BaRb$^+$ ion.
We probe the presence of a BaRb$^+$ ion by measuring whether a corresponding mass-filter removes the ion from the Paul trap, see Appendix \ref{sec:Massfilter}.
  The photodissociation laser is switched on $\tau=10\:\text{ms}$ after immersing the Ba$^+$ ion into the atom cloud without ODT. The observed decay of the BaRb$^+$ fraction can be approximately described by an exponential plus offset, $P_0 \exp(-\Gamma \Delta t) + P_{\infty}$ (see solid lines in Fig.$\:$\ref{fig4}).  The offset $P_\infty$ may stem from  BaRb$^+$ ions in certain vibrational levels of the states $(1)^3\Sigma^+$ or $(X)^1\Sigma^+$ which happen to have rather small photodissociation cross sections. As mentioned before, we expect the vibrational distributions in both states to be quite broad. In addition, the PECs indicate that the photodissociation cross sections for both the states $(1)^3\Sigma^+$ and  $(2)^1\Sigma^+$ by $493\,\text{nm}$ and $650\,\text{nm}$ light might be extremely small, because of a missing  Condon point at short range for the relevant transitions \cite{Hall2011}, see also Appendix \ref{sec:theophotodisso}.

From the measured laser intensities of $I_{493} = (180 \pm 40)\:\text{mW}\,\text{cm}^{-2}$ and $I_{650} = (260\pm 50)\:\text{mW}\,\text{cm}^{-2}$, we can determine effective, average  photodissociation cross sections for the given $(X)^1\Sigma^+$ BaRb$^+$ molecule population distribution over the vibrational states, using $\sigma =\Gamma h\nu / I$.
 We obtain $\sigma_\text{493} = (1.2\pm 0.3)\times 10^{-17}\:\text{cm}^2$ and $\sigma_\text{650} = (1.0\pm 0.2)\times 10^{-17}\:\text{cm}^2$.

We now test whether a Ba$^+$ ion has been produced during photodissociation, see hollow diamonds in Fig.$\:$\ref{fig4}. For the detection of the Ba$^+$ ion the mass-filtering scheme is used to remove a possibly remaining BaRb$^+$ ion, before long laser cooling and subsequent fluorescence imaging are carried out. For light at 650$\:$nm we do not find any Ba$^+$ signal. This can be explained with the help of Fig.$\:$\ref{fig1}(b). The 650$\:$nm laser couples the $(X)^1\Sigma^+$ state essentially only to the $(3)^1\Sigma^+$ state which, however, dissociates into $\text{Rb}^+ +\text{Ba}$. In contrast, for light at 493$\:$nm the production of Ba$^+$ ions is expected, and indeed the loss of BaRb$^+$ signal in Fig.$\:$\ref{fig4}(a) directly correlates with an increase of Ba$^+$ signal. Furthermore, we observe that about half of the produced Ba$^+$ ions end up in the metastable state $5D_{5/2}$, since their signal is lost as soon as we switch off the 614$\:$nm repump laser. Besides serving as a consistency check, this measurement also demonstrates that single ground state BaRb$^+$ molecules can be detected with high efficiency via fluorescence imaging.

From the experimentally determined cross sections  we can estimate
that when applying fluorescence imaging  photodissociation of a $(X)^1\Sigma^+$ state molecule will on average result in a (hot) Ba$^+$ ion with a probability of about $70\%$, and in a Rb$^+$ ion with a probability of about $30\%$.

\section{Conclusions and outlook}
\label{sec:conclusion}

In conclusion, we have studied the evolution of a BaRb$^+$ molecule in a gas of ultracold Rb atoms. We find that due to the high predictive power of the theory for the collisional and radiative processes of the BaRb$^+$ molecule only a comparatively small amount of experimental input is necessary to qualitatively pin down the evolution of the molecular ion. In order to experimentally probe the current state of the ion we have developed novel methods which are based on the coordinated concatenation of mass spectrometry, controlled photodissociation, timing of atom-ion interaction, laser cooling, and fluorescence imaging. We find that while the molecular evolution is dominated by vibrational relaxation for the most weakly-bound levels,  radiative processes become increasingly important for more deeply-bound levels. Furthermore, our work shows how differently the molecules behave depending on their electronic state. The holistic view of the molecular evolution presented here, opens up many new perspectives for future experiments, as it lays out how to prepare and manipulate specific molecular states and how to probe them. In the future, it will be interesting to extend the work presented here to resolve the vibrational and rotational states of the BaRb$^+$ ion. This will allow for investigating collisional and radiative processes and reaction paths so that our understanding can be tested on the quantum level.
Some of the methods presented here are very general and can be directly adopted for studies of a broad
range of other atomic and molecular species. These can be, e.g. of interest for research  in astrochemistry where reaction chains
in the cold interstellar medium are investigated \cite{Larsson2012, Snow2008, Petrie2007, Smith1992}.

\section*{Acknowledgments}

This work was supported by the German Research Foundation (DFG, Deutsche Forschungsgemeinschaft) within SFB/TRR21 and by the FP7 Marie-Curie Inititial Training Network COMIQ (Grant ID 607491) of the European Commission. A. Mohammadi acknowledges support from DAAD. The Ulm team thanks Joschka Wolf for assistance in the lab.

\section{Density evolution of atom cloud}
\label{sec:app_CloudDensity}

Initially, the prepared $^{87}$Rb atom cloud consists of about $N=6\times 10^6$ atoms \cite{Note5} and has a temperature of $T=750\:\text{nK}$. It is confined in a crossed ODT using laser light at $1064\:\text{nm}$. One ODT beam has a power of $1.6\:\text{W}$ and a beam waist of $230\:\upmu\text{m}$ at the location of the atoms. The other one has a power of $2.1\:\text{W}$ and a waist of $96\:\upmu\text{m}$. Dipole trap frequencies are $(\omega_x,\omega_y,\omega_z)=2\pi\times(145,145,22)\:\text{Hz}$ for the three directions of space $i\in\{x,y,z\}$. Here, the $y$-axis corresponds to the vertical axis, which is along the direction of the acceleration of gravity $g$. The initial widths $\sigma_{i,0}$ of the atomic cloud are $\sigma_{i,0}= \omega_i^{-1}\sqrt{k_\text{B}T/m_\text{Rb}}$, where $m_\text{Rb}$ is the atomic mass of $^{87}$Rb. We obtain $\sigma_{x,0}=\sigma_{y,0}=9\:\upmu \text{m}$, and $\sigma_{z,0}=60\:\upmu \text{m}$.

When switching off the ODT at a time $t=0$, the evolution of the density $n(t)$ of the atomic cloud at the position of the ion can be expressed by
\begin{equation}
	n(t)=\frac{N(2\pi)^{-3/2}}{\sigma_x(t)\sigma_y(t)\sigma_z(t)}\exp{\left(-\frac{g^2 t^4}{8\sigma_y^2(t)}\right)\,,}
	\label{eq:desnityevl}
\end{equation}
using the atom cloud widths $\sigma_i(t)=\sqrt{\sigma_{i,0}^2+k_\text{B}Tt^2/m_\text{Rb}}$. In Fig.$\:$\ref{fig2}(b) we show the density evolution of the atom cloud at the location of the ion trap center. We note that the interaction time $\tau$ is given by $\tau=t-250\:\upmu\text{s}$, since the atoms are released $250\:\upmu \text{s}$ before the ion is immersed into the atom cloud at $\tau=0$. In Fig.$\:$\ref{fig3} we have essentially the same density evolution despite the fact that at $\tau'$ some laser light at $1064\:\text{nm}$ is switched on. We have checked numerically that due to the low intensity of $1.8\:\text{kW}\,\text{cm}^{-2}$ used for these measurements the effect of the optical trapping potential is negligible.

\section{Mass filtering}
\label{sec:Massfilter}

We can selectively remove an ion of a pre-chosen mass from the Paul trap by resonantly heating the ion out of the trap. For this we modulate the voltages on
electrodes which are normally used for the compensation of radial stray electric fields at the ion trap center
\cite{Haerter2013a,MohAplPhysB2019}.
This modulation shifts the trap center periodically about the axial symmetry axis of the Paul trap. The frequency of the modulation is set to be the  mass-dependent trap frequency of the chosen ion species.  We typically modulate the trap for a duration of 3$\:$s.
We have performed test measurements for deterministically prepared Ba$^+$, Rb$^+$, and Rb$_2^+$ ions. In these cases we observed an efficiency of almost 100\% for removing the ion by resonant modulation. We therefore also expect a similar efficiency for a BaRb$^+$ ion. A modulation with the resonance frequency for a particular ion species does not affect the trapping of an ion of a different species relevant for the present work.

\section{Detection of the ion}
\label{sec:app_Detection_ion}

In order to detect a single, trapped ion in the Paul trap we have two methods which we describe in the following.

\subsection{Fluorescence detection of a single Ba$^+$ ion}
\label{sec:Fluores_img}

In order to detect a Ba$^+$ ion we first separate the ion trap center from the atom trap center by a distance of $100\:\upmu$m which is much larger than the size of the atomic cloud in order to suppress unwanted collisions. This is done by applying appropriate dc voltages on the Paul trap endcap electrodes. Afterwards, the atoms are released from the ODT by switching it off. After $20\:$ms, when all atoms have left, we move the ion back to its former position, since this position corresponds to the centers of the cooling laser beams for the Ba$^+$ ion.  Here, the lasers have beam waists ($1/e^2$ radii) of about 20$\:\upmu$m. The cooling laser beams consist of one beam at a wavelength of 493$\:$nm for driving the $6S_{1/2}$ to $6P_{1/2}$ Doppler cooling transition, and one beam at a wavelength of 650$\:$nm for repumping the Ba$^+$ ion from the metastable $5D_{3/2}$ state towards $6P_{1/2}$.  During a laser-cooling time of $100\:\text{ms}$ an electron multiplying CCD camera takes a first fluorescence image of the Ba$^+$ ion. This method allows for detection of a cold Ba$^+$ ion with a temperature of $T \approx100\:\text{mK}$ or below, due to the short duration of the laser cooling.  A hotter Ba$^+$ ion, e.g. resulting from photodissociation with a kinetic energy on the order of $0.2\:\text{eV}$, can be detected by taking a second image after long laser cooling. For this, the 493$\:$nm laser beam frequency is red-detuned by $1\:\text{GHz}$ and swept back towards resonance within three seconds.  Afterwards, we take another fluorescence image, again for a duration of $100\:\text{ms}$.  From the two images we can discriminate a hot ion from a cold one.  For example, if a fluorescing Ba$^+$ ion is found in the second image but not in the first one, then this Ba$^+$ ion was hot at the time of the first image.
 Furthermore, we can detect whether a Ba$^+$ ion is in the metastable state $5 D_{5/2}$. Such an ion will only  appear in the fluorescence image, if we previously pump it out of the $5D_{5/2}$ state, e.g. with a 614$\:$nm laser via the $6P_{3/2}$ level.
Therefore, in order to probe for a $5D_{5/2}$ ion, we take two sets of fluorescence images. The first set is without the 614$\:$nm repump laser and the second set is with the 614$\:$nm repump laser. If we only obtain a fluorescence signal in the second set of the images, then the Ba$^+$ ion was in the metastable state $5 D_{5/2}$.

\subsection{Detection of the ion via atom loss and discrimination of ion species}
\label{sec:detect2}

In our setup only the Ba$^+$ ion can be detected directly via fluorescence imaging. In order to detect a different ion species such as BaRb$^+$, Rb$^+$, Rb$_2^+$ we use a scheme where the ion inflicts atom loss in a cold atom cloud \cite{Haerter2013a, Haerter2013}.
 For this, the ion is kept in the ion trap while a new cloud of neutral atoms is prepared. Then, the ion is immersed into this new atom cloud.
  Elastic collisions of the ion with the ultracold atoms lead to loss of atoms as they are kicked out of the ODT, which is much shallower than the ion trap. After a given interaction time the remaining number of atoms is measured via absorption imaging. If this number is significantly lower than for a reference measurement using a pure atom cloud  an ion is present.
Typically we already know from the preparation procedure (and, because all relevant ion species cannot escape from the deep Paul trap potential), that a single ion must be trapped in the Paul trap, but we would like to discriminate between the ion species BaRb$^+$, Rb$^+$, and Rb$_2^+$.
  For this, we carry out mass-filtering in the Paul trap (see Appendix \ref{sec:Massfilter}), where we remove selectively the ion from the trap if it has a
  specific, pre-chosen mass. Subsequently, we test whether the ion has been removed from the Paul trap with the ion detection scheme based on inflicted atom loss.

\section{Calculation of cross sections}
\label{sec:app_ColCrossSec}

\subsection{Cross sections from QCT calculations}
\label{sec:QTC}

\subsubsection*{Model}

We use QCT calculations \cite{Perez2019} to determine cross sections for elastic collisions, vibrational relaxation, collisional dissociation and substitution reactions in collisions
of a BaRb$^+$ ion with an ultracold Rb atom.

Since the three-body process occurs at large internuclear distances we assume that the three-body potential energy surface can be described by pair-wise additive ground-state potentials according to $V(\vec{R}_1,\vec{R}_2,\vec{R}_3)=V(\vec{R}_{12})+V(\vec{R}_{13})+V(\vec{R}_{23})$. Here, the Rb-Rb interaction is taken from \cite{Strauss2010}, while the Ba$^+$-Rb and Ba-Rb$^+$ interactions are modeled by means of the generalized Lennard-Jones potential $V(R)=-C_4[1-(R_m/R)^4/2]/R^4$, where
$C_4 = 160\:\text{a.u.}$, $R$ is the internuclear distance, $R_m=9.27\:a_0$, and $a_0$ is the Bohr radius.
  We note in passing that $C_4 = \alpha_\text{Rb}e^2/[2(4\pi\varepsilon_0)^2]$ is proportional to the
static dipolar polarizability $\alpha_\text{Rb} = 4\pi\varepsilon_0 \times 4.739(8) \times 10^{-29}$m$^3$ of the Rb atom \cite{gregoire2015}.
 $e$ is the elementary charge and $\varepsilon_0$ is the vacuum permittivity.   The Lennard-Jones type potential describes the long-range interaction correctly and leads to a manageable computational time.
We note that the $C_4$ coefficient for a Ba atom is $134\pm10.8\:$a.u. \cite{Shafer2007}. Nevertheless, for saving computational time we simply use the same coefficient for the Ba atom as for the Rb atom in our model. This introduces small quantitative errors of about 5\% but does not change the qualitative interpretation.

Using the QCT approach, we study the collisional behavior of BaRb$^+$ molecules in the states $(2)^1\Sigma^+$,  $(1)^3\Sigma^+$ as well as $(X)^1\Sigma^+$.  We ignore any spin degrees of freedom, which means that the results are the same for both $(2)^1\Sigma^+$ and $(1)^3\Sigma^+$ BaRb$^+$ molecules.
 Furthermore, we only consider collisions where
BaRb$^+$ molecules are initially nonrotating, i.e. $j=0$.
We have numerically checked that for other low $j$-states the results will not be significantly different at the level of our approximations.
In order to determine cross sections and rates for a given electronic and vibrational state we sum over the corresponding rotational distribution of the final products.

\begin{figure}[t]
\centering\includegraphics[width=0.8\columnwidth]{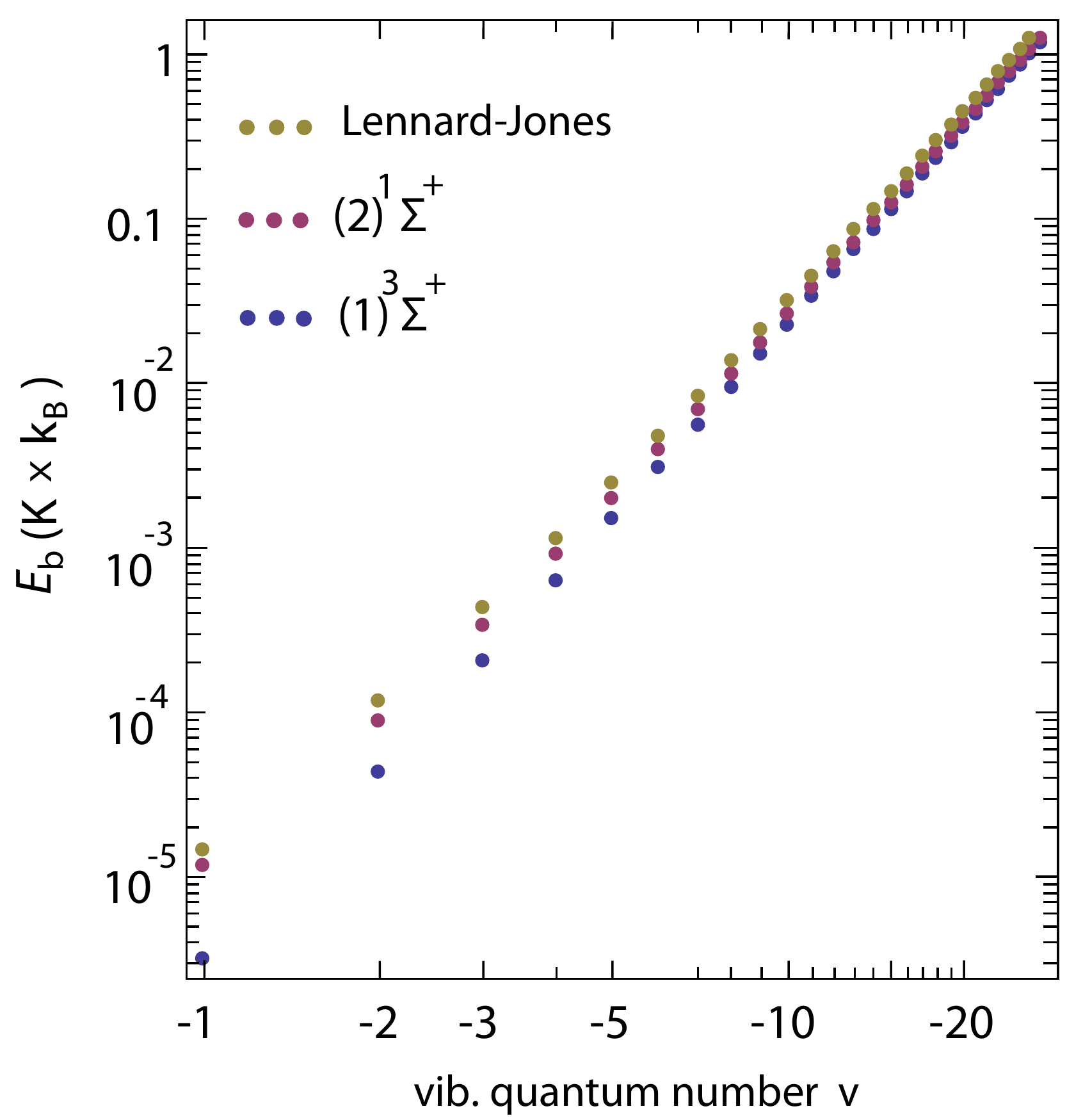}
\caption{Vibrational binding energies of a BaRb$^+$ molecule for the electronic states $(1)^3\Sigma^+$ and $(2)^1\Sigma^+$.
 The rotation is in the ground state, i.e. $j = 0$.
  The brown data points result from the  Lennard-Jones potential which is used for the QCT calculations.
  The purple and blue data points are the results from our calculated  $(1)^3\Sigma^+$ and $(2)^1\Sigma^+$ PECs, as described in Appendix \ref{sec:PECs}.}
\label{fig:Ebinds}
\end{figure}

 In Fig.$\:$\ref{fig:Ebinds}, the energetically uppermost vibrational levels as derived from the Lennard-Jones potential are shown down to binding energies of about 1$\:$K$\times k_\text{B}$.  For comparison, we also present the results from the PEC calculations
 for the $(2)^1\Sigma^+$ and $(1)^3\Sigma^+$ electronic states (see Appendix \ref{sec:PhotoDisso}).

\begin{figure}[t]
\centering\includegraphics[width=\columnwidth]{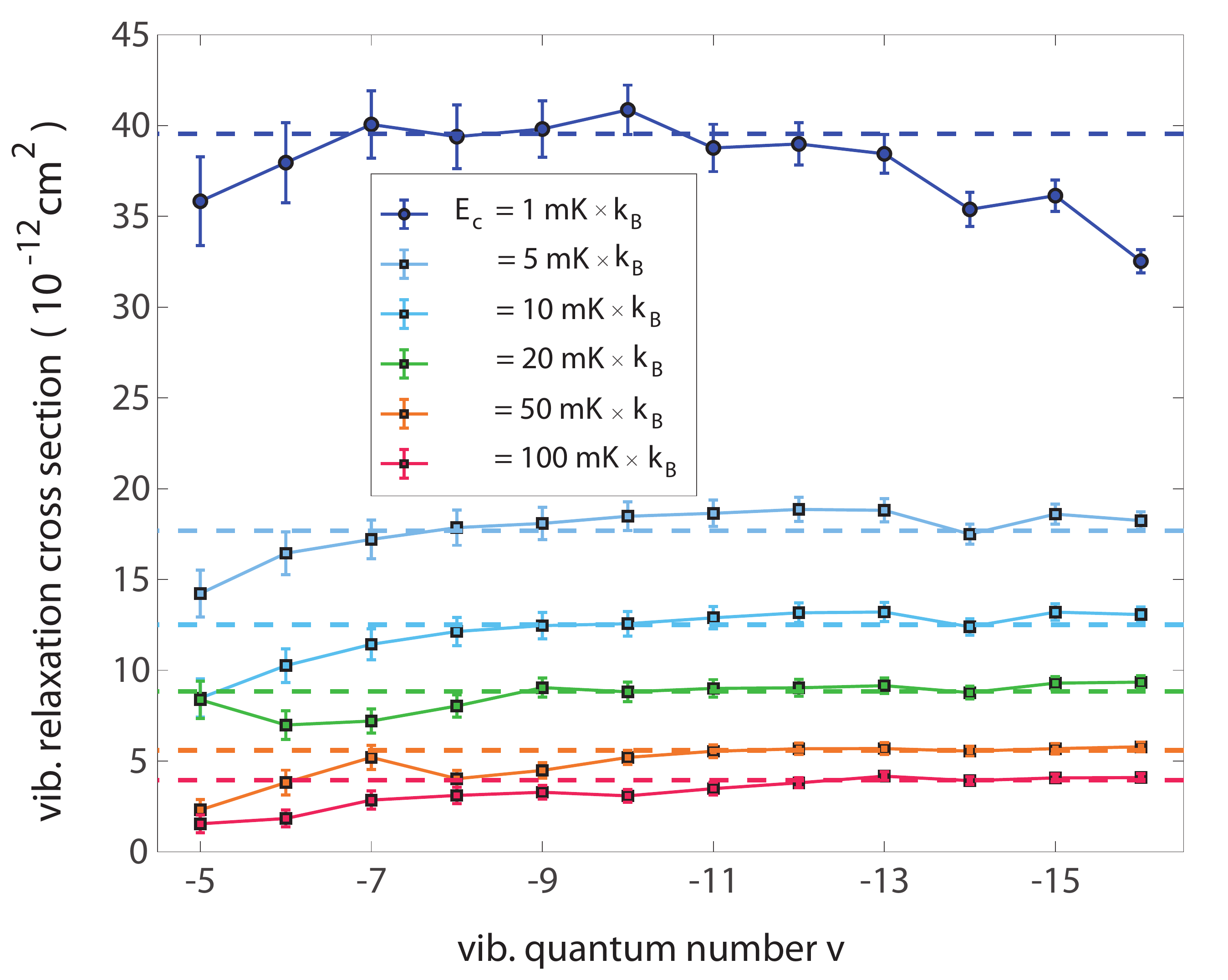}
\caption{Vibrational relaxation cross section as a function of the vibrational quantum number $v$, for various collision energies $E_\text{c}$. Shown are results of QCT calculations. The error bars represent $1\sigma$ standard deviation obtained by evaluating the numbers of trajectories leading to the same outcomes. The dashed lines are the Langevin cross sections.}	
\label{fig:Vib_qunch}
\end{figure}

We have carried out QCT calculations for the vibrational levels $v=(-1,-2,...,-16)$ and for a collisional energy range of $E_\text{c}=1-100\:$mK$\times k_\text{B}$.
 This range for $E_c$ corresponds to a range of the
  kinetic energy of the BaRb$^+$ ion of $E_\text{c}(1-\mu/m_\text{Rb})^{-1}=3.6-360\,\text{mK}\times k_\text{B}$, when assuming zero kinetic energy for the atoms. Here, $\mu$ is the reduced mass of the Rb-BaRb$^+$ system. For a given set of $v$ and $E_\text{c}$, we determine a suitable maximum impact parameter $b_\text{max}$ beyond which no reactions/inelastic processes occur anymore. $b_\text{max}$ is typically on the order of the Langevin radius $b_\text{L}=(4C_4/E_c)^{1/4}$.  We run batches of 10$^4$ trajectories, effectively sampling the configuration space including different impact parameters $b < b_\text{max}$ and molecular orientations. As a result we obtain a probability distribution for the different collisional processes. The cross section for a specific collision process $\kappa$  can be calculated as $\sigma_\kappa = \pi \, b_\text{max}^2 \, P_\kappa$, where $P_\kappa$ is the probability for a trajectory undergoing this process.

\subsubsection*{Results}

\begin{itemize}
\item  \textit{Vibrational relaxation:} Figure \ref{fig:Vib_qunch} shows the cross sections for vibrational relaxation for the states $(2)^1\Sigma^+$ and  $(1)^3\Sigma^+$ for different collision energies. The calculations clearly reveal that in general the vibrational relaxation cross section is well approximated by the Langevin cross section $\sigma_{\text{L}}(E_\text{c})= \pi \sqrt{4C_4/E_\text{c} }$. The corresponding Langevin rate $\Gamma_\text{L}(t)=\sigma_{\text{L}}v_\text{ion}n(t)=K_\text{L}n(t)$ is independent of the collision energy. Here, $v_\text{ion}=\sqrt{2E_\text{c}/\mu}$ is the velocity of the BaRb$^+$ ion and $K_\text{L} =2\pi \sqrt{2C_4/\mu}=2.03\times10^{-9}\:\text{cm}^{3}\text{s}^{-1}$ is the Langevin rate constant. We note that in our calculations vibrational relaxation typically leads to a change in the vibrational quantum number $v$ by one or two units, i.e. $v'=v-1, v-2$.  The average change is 1.4 units, as discussed later in Appendix \ref{subs:Evolution}. Since these results are quite independent of the initial vibrational quantum number (see Fig.\ref{fig:Vib_qunch}), we adopt them for levels which are more deeply-bound than $v = -16$. Furthermore, we use them also for vibrational relaxation in the ground state $(X)^1\Sigma^+$.  If the collision energy is large enough, in principle, also vibrational excitation could occur, but for our settings the calculations show that this is quite negligible.

\item \textit{Substitution reaction} $\text{BaRb}^+ + \text{Rb} \rightarrow \text{Rb}_2 + \text{Ba}^+$: For the weakly-bound levels of the $(2)^1\Sigma^+$ and $(1)^3\Sigma^+$ electronic states this reaction is in general so rare that it can be neglected.  This can be explained as follows in a simple classical picture. The Ba$^+$  ion and the Rb atom of the weakly-bound BaRb$^+$ molecule are generally well separated. The colliding free Rb atom mainly interacts with the Ba$^+$ ion via the long-range polarization potential while the interaction between the two Rb atoms is essentially negligible. Hence, the formation of the neutral Rb$_2$ molecule is unlikely. For the ground state $(X)^1\Sigma^+$ the reaction is energetically closed.

\item \textit{Substitution reaction} $\text{BaRb}^++ \text{Rb} \rightarrow \text{BaRb} + \text{Rb}^+$: For the weakly-bound levels of the $(2)^1\Sigma^+$, $(1)^3\Sigma^+$ and $(X)^1\Sigma^+$ electronic states this reaction is  rare because it entails the formation of a neutral molecule, following similar arguments as for the previously discussed reaction BaRb$^+$ + Rb $\rightarrow$ Rb$_2$ + Ba$^+$. In addition, for the weakly-bound levels of $(2)^1\Sigma^+$ and $(1)^3\Sigma^+$ the reaction would require a charge transfer between the Rb atom and the Ba$^+$ ion, since for these electronic states and long binding lengths the positive charge is almost completely located on the Ba atom within the BaRb$^+$ molecule.
For deeply-bound levels in the ground state $(X)^1\Sigma^+$ with $v \lesssim 90$
the reaction is energetically closed. This covers about 70$\%$ of the produced ground state molecules, as discussed in Appendix \ref{sec:AppendRadRelax}. For these reasons we ignore this substitution reaction in our model.

\item  \textit{Substitution reaction} $\text{BaRb}^+ +\text{Rb} \rightarrow \text{Rb}_2^+ + \text{Ba}$: For the weakly-bound levels of the states $(2)^1\Sigma^+$ and $(1)^3\Sigma^+$ this reaction involves a charge exchange and is therefore negligible. For the weakly-bound levels of the ground state $(X)^1\Sigma^+$, where the positive charge of the  BaRb$^+$ molecule is located on the Rb atom,
the substitution reaction can have a sizable probability. From numerical QCT calculations for the most weakly-bound levels we can extrapolate roughly the scaling law $\sigma \approx \, a_0^2 \, E_b / (\text{mK} \times k_B)$ for the cross section.
Thus, the cross section increases linearly with the binding energy. We expect this to be approximately valid up to a binding energy of about 1000$\:\text{K}\times k_\text{B}$, where the expression  should smoothly go over to the Langevin cross section.

\item \textit{Elastic collisions:} Due to the restriction $b < b_\text{max}$ we do in general not take into account all elastic collisions. In particular those with very little energy transfer are omitted since
     they are irrelevant for sympathetic cooling. To a first approximation, the elastic cross section for which sizable amounts of kinetic energy are transferred between the collision partners is the Langevin cross section.  This is valid for
    all states, i.e. $(2)^1\Sigma^+$, $(1)^3\Sigma^+$, and $(X)^1\Sigma^+$.

\item \textit{Collisional dissociation:} A BaRb$^+$ molecule can dissociate in a collision with a Rb atom, if the collision energy $E_\text{c}$ is large enough. For a weakly-bound
BaRb$^+$ molecule in the state $(1)^3\Sigma^+$ or $(2)^1\Sigma^+$
this would lead to a release of a Ba$^+$ ion and a Rb atom.
In our experiments, however, the typical collision energy is too small. Therefore, this process is relevant only for the most weakly-bound BaRb$^+$ molecules. In our simulation this only concerns the vibrational level $v=-5$  in the states $(2)^1\Sigma^+$ and $(1)^3\Sigma^+$. It has  a binding energy of $\approx 2\:\text{mK}\times k_\text{B}$.
For a collision energy of $E_\text{c}=2\:\text{mK}\times k_\text{B}$, which is typical in the beginning of our experiments, the calculated dissociation cross section is then $4\times10^{-12}\,\text{cm}^{2}$ \cite{Perez2019}. This is about a factor of seven smaller than the Langevin cross section.

\end{itemize}

\subsection{Spin-flip cross section}
\label{ssec:spin-flip}

If the BaRb$^+$ ion is in the triplet $(1)^3\Sigma^+$ (singlet $(2)^1\Sigma^+$) state it may undergo an electronic spin-flip towards the singlet $(2)^1\Sigma^+$ (triplet $(1)^3\Sigma^+$) state in a close-range collision with a Rb atom. Discussions of spin-flip processes for molecules can be found in the literature, see, e.g., \cite{Krems2004,Avdeenkov2001,Krems2008}.

We estimate the spin-flip cross section in the following way. In the collision between a Rb atom and a BaRb$^+$ molecular ion we only consider the interaction between the free Rb atom and the Ba$^+$ ion which is loosely bound in the BaRb$^+$ molecule.
Spin-flips can occur when the two electron spins of the Ba$^+$ ion and the free Rb atom are opposite to each other, e.g. $m_s(\text{Ba}^+) = 1/2$ and $m_s(\text{Rb})=-1/2$, such that after the collision the spins are $m_s(\text{Ba}^+) = -1/2$ and $m_s(\text{Rb})=1/2$. Here, $m_s$ is the magnetic quantum number of the electron spin.
 Taken by itself, the state $m_s(\text{Ba}^+) = 1/2$, $m_s(\text{Rb})=-1/2$ is
 a 50\% / 50\% superposition state of spin singlet and spin triplet. In the following we  estimate the spin-flip cross section for such a superposition state. The actual spin-flip cross section for our experiment should be a fraction of this, because the statistical factors of the total spin-decomposition need to be taken into account. This requires an analysis, in how far a spin-flip of the bound Ba$^+$ ion leads to a flip of the total electron spin in the BaRb$^+$ molecule. Such an analysis is, however, beyond the scope of the present work.
 
 The spin-exchange cross section for the 50\% / 50\% superposition state can be estimated using a partial-wave approach \cite{Dalgarno1965, Cote2000} as
\begin{equation}
\label{spin-exchange}
\sigma_\text{sf}(E_{c})=\frac{\pi}{k^2}\sum_{l}(2l+1)\sin^2{\left(\delta_{l}^{S}(E_\text{c})-\delta_{l}^{T}(E_\text{c})\right)}\,,
\end{equation}
where $\delta_{l}^{S}(E_\text{c})$ and $\delta_{l}^{T}(E_\text{c})$ are the energy dependent phase-shifts of the partial wave $l$  for the singlet and  triplet atom-ion potential energy curves, respectively. Here, $k$ is the wave number of the relative momentum in the center-of-mass frame. Next, we determine an angular momentum $l_\text{max}$ such that for $l>l_\text{max}$ the phase-shift $\delta_{l}^{S}(E_\text{c})\approx\delta_{l}^{T}(E_\text{c})$. This is possible, because for large enough $l$ the particles only probe the long-range tail  of the ion-atom potential and this tail is essentially the same for singlet and triplet states. Therefore, only trajectories with $l\le l_{\text{max}}$ contribute to the cross section. For $l\le l_{\text{max}}$ we estimate the contribution of each partial wave term in Eq.$\:$(\ref{spin-exchange}) by using the random phase approximation for the phase-shifts, $\sin^2{\left(\delta_{l}^{S}(E_\text{c})-\delta_{l}^{T}(E_\text{c})\right)} = 1/2$~\cite{Cote2000}. The partial wave $l_\text{max}$ can be estimated \cite{Ratschbacher13} using  the critical impact parameter (Langevin radius) via $l_\text{max} =  b_\text{L} k  = (2C_4/E_\text{c})^{1/4} k$, as for impact parameters $b> b_\text{L}$ the inelastic cross section vanishes in the classical regime. Carrying out the sum in Eq.$\:$(\ref{spin-exchange}) up to $l_\text{max}$ we obtain
\begin{equation}
\label{Sigma_ex}
\sigma_{\text{sf}}(E_\text{c})=\frac{\pi l_{\text{max}}^2}{2k^2} =  {\pi \over 2} \left( {2 C_4\over E_\text{c}} \right)^{1/2} = {\sigma_{\text{L}}(E_\text{c})\over 2}\,.
\end{equation}
The spin-flip rate is then simply proportional to the Langevin rate $\Gamma_{\text{L}}$. We stress again, that Eq.$\:$(\ref{Sigma_ex}) is only an estimate. Therefore, for our simulations we allow for another constant fit parameter $\gamma$ such that the spin-flip rate for a transition from the singlet state to the triplet state is given by
\begin{equation}
\label{eq:gamma_Adp_factor}
\Gamma_\text{sf} = \gamma \,  \Gamma_{\text{L}} \, .
\end{equation}
For determining $\gamma$ we use experimental data for which the ODT is off. When setting  $\gamma =  1/ 14$ we obtain good agreement with our measurements.

Finally, we note that the spin-flip rate for a transition from a triplet state to a singlet state is not $\Gamma_\text{sf}$ but $\Gamma_\text{sf}/3$. This is because a spin-flip process between the bound Ba$^+$ ion and the free Rb atom does not necessarily change a triplet BaRb$^+$ molecule into a singlet one. Collisions with an unpolarized sample of Rb will in general shuffle around the total spin $S$ of the molecule equally between the four levels $S = 0, m_S = 0$ and $S = 1, m_S = -1, 0, 1$. Thus, on average only 1 in 3 spin-exchange collisions of a triplet BaRb$^+$ molecule will produce a singlet BaRb$^+$ molecule.

\subsection{Radiative relaxation and photodissociation cross sections}
\label{sec:PhotoDisso}

In the following we calculate cross sections for radiative relaxation and photodissociation.
For this, we first calculate PECs, wave functions and transition dipole moments. Table \ref{tab:ListOfLevels} shows electronic states for relevant transitions. In the following discussion, the spin-orbit interaction will be neglected.

\begin{table}
\caption{List of the relevant electronic states, $\Lambda$, for the BaRb$^{+}$ molecule that can be reached from the entrance channel upon absorption of a photon of 1064$\:$nm wavelength.  The entrance channel is spanned by the electronic states $(2)^1\Sigma^+$ and $(1)^3\Sigma^+$ which correlate in the asymptotic limit to
$\text{Rb}\left(5s\,^{2}S\right)+\text{Ba}^{+}\left(6s\,^{2}S\right)$.   $E_\text{k,max}$ is the atom-ion relative kinetic energy released after photodissociation when the initial molecule is weakly bound. $\sigma_\text{max}$ is the largest estimate for the state-to-state absorption cross section for a BaRb$^{+}$ molecule with binding energy $E_b = 1\:\text{K}\times k_\text{B}$
 (see also Appendix \ref{sec:theophotodisso}).
}
\label{tab:ListOfLevels}
\begin{ruledtabular}
\begin{tabular}{llll}
$\Lambda$ & Asymptotic limit &$E_\text{k,max}$& $\sigma_\text{max}$ \\
 & &(cm$^{-1}$)&(cm$^{2}$)  \\
\hline
 $\left(2\right)^{3}\Sigma^{+}$  &  $\text{Rb}^{+}+\text{Ba}\left(6s5d\,^{3}D\right)$  & $\approx8200$  & negligible \tabularnewline
 $\left(1\right)^{3}\Pi$  &  &  & negligible\tabularnewline
&&&\\
 $\left(3\right)^{1}\Sigma^{+}$  &  $\text{Rb}^{+}+\text{Ba}\left(6s5d\,^{1}D\right)$  &  $\approx5900$  & $\approx10^{-25}$\tabularnewline
  $\left(1\right)^{1}\Pi$  &  &  & $\approx10^{-27}$\tabularnewline
&&&\\
  $\left(3\right)^{3}\Sigma^{+}$  &  $\text{Rb}^{+}+\text{Ba}\left(6s6p\,^{3}P\right)$ & $\approx4800$ &  $\approx 9 \times 10^{-20}$\tabularnewline
  $\left(2\right)^{3}\Pi$  &  &  &  $\approx 4 \times 10^{-20}$\tabularnewline
&&&\\
  $\left(4\right)^{1}\Sigma^{+}$  &  $\text{Rb}\left(5s\,^{2}S\right)+\text{Ba}^{+}\left(5d\,^{2}D\right)$  & $\approx4000$  &  $\approx 4 \times 10^{-19}$\tabularnewline
  $\left(2\right)^{1}\Pi$  &  &  & $\approx 2 \times 10^{-27}$\tabularnewline
  $\left(4\right)^{3}\Sigma^{+}$  &  &  & $\approx 4 \times 10^{-21}$\tabularnewline
  $\left(3\right)^{3}\Pi$  &  &  &  $\approx 2 \times 10^{-21}$\tabularnewline
\end{tabular}
\end{ruledtabular}
\end{table}

\subsubsection*{Potential energy curves}
\label{sec:PECs}

The PECs displayed in Fig.$\:$\ref{fig1}(b), the permanent electric dipole moments (PEDMs), and the transition electric dipole moments (TEDMs) for the BaRb$^{+}$ molecule are obtained by the methodology described, e.g., in \cite{aymar2005,aymar2006a,aymar2011}. Briefly, the calculations are carried out using the Configuration Interaction by Perturbation of a Multiconfiguration Wave Function Selected Iteratively (CIPSI) package \cite{huron1973}. The electronic structure is modeled as an effective system with two valence electrons moving in the field of the Rb$^{+}$ and Ba$^{2+}$ ions represented by effective core potentials (ECP), including relativistic scalar effects, taken from Refs.$\:$\cite{durand1974,durand1975} for Rb$^{+}$ and Refs.$\:$\cite{fuentealba1985,fuentealba1987} for Ba$^{2+}$. The ECPs are complemented with core polarization potentials (CPP) depending on the orbital angular momentum of the valence electron \cite{muller1984a,foucrault1992}, and parametrized with the Rb$^{+}$ and Ba$^{2+}$ static dipole polarizabilities and two sets of three cut-off radii \cite{guerout2010,aymar2012}. Only the remaining two valence electrons are used to calculate the Hartree-Fock and the excitation determinants, in atom-centered Gaussian basis sets, through the usual self-consistent field (SCF) methodology. The basis set used for the Rb atoms is from Refs.$\:$\cite{aymar2005,aymar2006a}, and the one for Ba is from Refs.$\:$\cite{aymar2012,bouissou2010}. A full configuration interaction (FCI) is finally achieved to obtain all relevant PECs, PEDMs, and TEDMs. In Ref.$\:$\cite{Silva2015} a comparison between these calculations for several systems, including BaRb$^{+}$, and the ones available in the literature is given for the $(X)^{1}\Sigma^{+}$
and (2)$^{1}\Sigma^{+}$ electronic states.

Since for the states $(1)^{3}\Sigma^{+}$ and $(2)^{1}\Sigma^{+}$ we need to consider extremely
weakly-bound vibrational levels, PECs have to be calculated up to large inter-particle distances.
For this, we analytically extend the existing short-range PECs by matching them to the atom-ion long-range interaction behavior
\begin{equation}
\lim_{R\rightarrow\infty}V\left(\Lambda;\,R\right)=D_{e}-\frac{C_{4}}{R^{4}}\,,
\label{eq:LongRangeDef}
\end{equation}
where $D_{e}$ is the dissociation energy of the electronic state $\Lambda$. From fits of Eq.$\:($\ref{eq:LongRangeDef}) to our \textit{ab initio} PECs at around $25\:a_0$, we obtain a $C_{4}$ value of about $C_{4} = 171\:\text{a.u.}$, which is close to the known value $C_{4}=160\:\text{a.u.}$ for Rb atoms.

We note that the asymptotic energies ($R \rightarrow \infty$) for our PECs are in reasonable agreement with experimental values. There is virtually no error regarding those asymptotes for which each valence electron is localized on one atomic core [e.g., for the asymptotes $\text{Rb}\left(5s\,^{2}S\right)+\text{Ba}^{+}\left(6s\,^{2}S\right)$ and $\text{Rb}\left(5s\,^{2}S\right)+\text{Ba}^{+}\left(5d\,^{2}D\right)$]. However, if both valence electrons are localized on the Ba atom we obtain deviations from experimental values of $-180\:\text{cm}^{-1}$ for the $\text{Rb}^{+}+\text{Ba}\left(6s^{2}\,^{1}S\right)$ asymptote, $-120\:\text{cm}^{-1}$ for the $\text{Rb}^{+}+\text{Ba}\left(6s6p\,^{3}P\right)$ asymptote, and $420\:\text{cm}^{-1}$ for the $\text{Rb}^{+}+\text{Ba}\left(6s5d\,^{1}D\right)$ asymptote, respectively \cite{aymar2012}.

\subsubsection*{Calculation of wave functions}

The diatomic eigenvalue problem is solved for each PEC $V\left(\Lambda;\,R\right)$ by means of the mapped Fourier grid Hamiltonian (MFGH) method \cite{kokoouline1999}, which diagonalizes a discrete variable representation (DVR) matrix of the Hamiltonian. We use a fairly large internuclear distance range, $R_{\mathrm{max}}\approx5000\:a_{0}$, in order to even accommodate small binding energies $E_{b}$ on the order of $E_{b}\approx 10\,\upmu\text{K}\times k_\text{B}$.

The energy-normalized continuum wave functions $\left|\Lambda^{\prime}j^{\prime};\,k\right\rangle $ are computed using a standard Numerov method \cite{johnson1978}. Here, $j^{\prime}$ is the rotational quantum number. Since the kinetic energies at long range for the exit channels correspond to several thousands of wave numbers the calculations are performed on a fairly dense and large grid (between 90,000 and 150,000 grid points) so that there are at least 20 points per wave function oscillation.

\subsubsection*{Transition electric dipole moments}

\begin{figure}[t]
\centering\includegraphics[width=1\columnwidth]{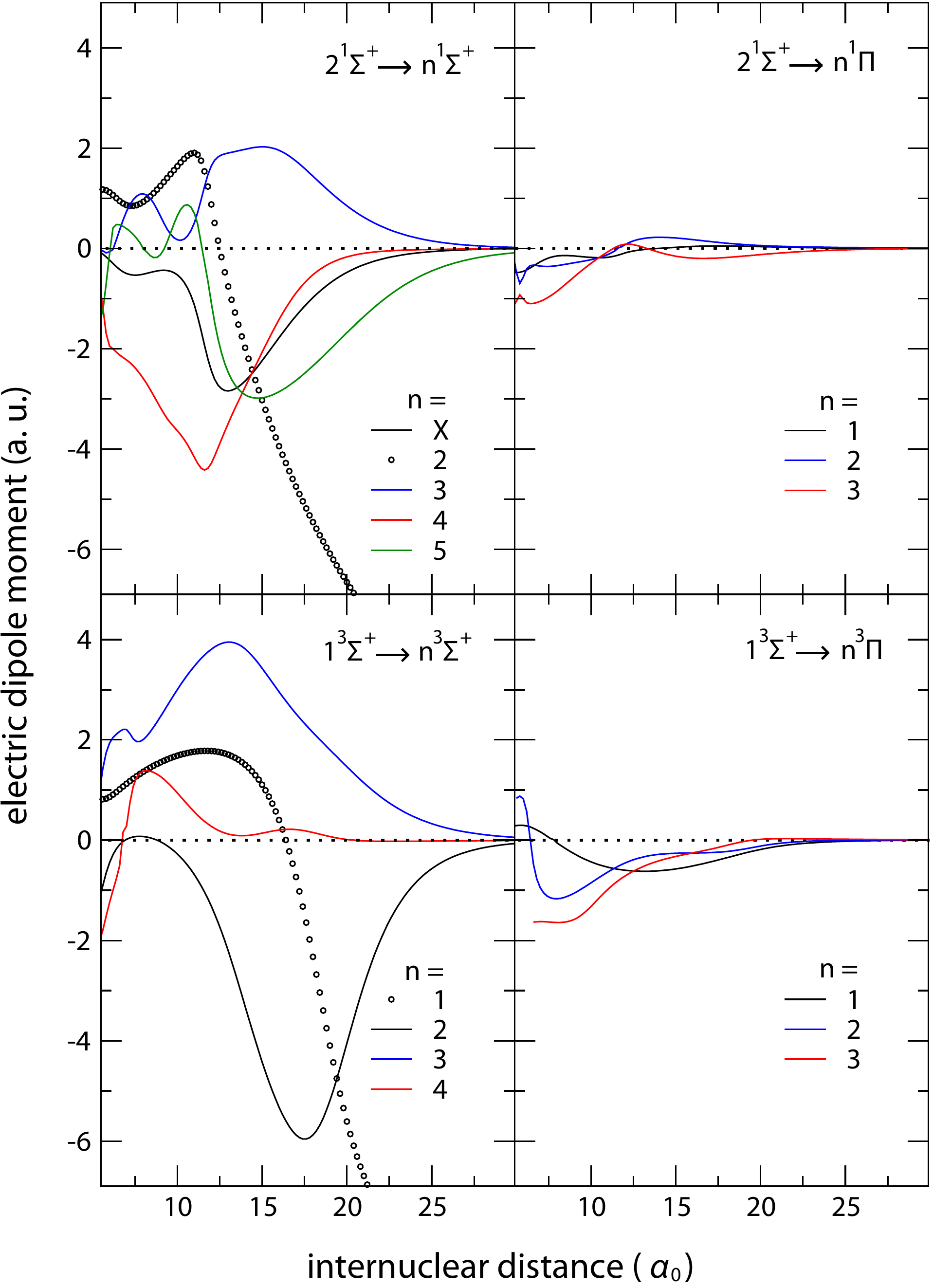}
\caption{PEDMs (circles) and TEDMs (solid lines) as functions of the internuclear distance of the BaRb$^{+}$ molecule. Initial and final states are given in the plot. For $\Sigma$--$\Sigma$ transitions the dipole moment along the internuclear axis is shown, whereas for $\Sigma$--$\Pi$ transitions the dipole moment in transverse direction is shown.}
\label{fig:TEDM}
\end{figure}

The TEDMs $D_{\Lambda^{\prime},\Lambda}\left(R\right)$ between relevant electronic states $\Lambda$ and $\Lambda^{\prime}$ are shown in Fig.$\:$\ref{fig:TEDM} as functions of the internuclear distance $R$. The plots show that $\Sigma$--$\Sigma$ transitions are generally stronger than $\Sigma$--$\Pi$ transitions. Furthermore, the TEDMs vanish at large distances. Such a behavior is expected, since the $\text{Ba}^++\text{Rb}$ asymptote cannot be addressed from the $\text{Rb}^++\text{Ba}$ asymptote by optically exciting one of the atoms [see Fig.$\:$\ref{fig1}(b)]. Therefore, for weakly-bound rovibrational states all outer turning points can be disregarded, i.e. radiative processes are driven at short range.

\subsubsection*{Photodissociation}
\label{sec:theophotodisso}
In order to determine the photodissociation cross sections, we calculate the absorption cross sections $\sigma_{\Lambda^{\prime}j^{\prime}k,\,\Lambda\upsilon j}\left(E^{\prime}\right)$ for the transitions between rovibrational levels $(v, j)$ in the electronic state $\Lambda$ towards the continuum of an electronic state $\Lambda'$ \cite{kirby1989,bovino2011c}
\begin{equation}
\begin{split}
\label{eq:StateToStateCrossSectionDef}
\sigma_{\Lambda^{\prime}j^{\prime}k,\,\Lambda\upsilon j}&\left(E^{\prime}\right)= \\
& \frac{4\pi^{2}}{3c}\frac{E^{\prime}}{2j+1}S\left(j^{\prime},j\right)\left|\left\langle \Lambda^{\prime};\,k\right|D_{\Lambda^{\prime},\Lambda}\left(R\right)\left|\Lambda\upsilon \right\rangle \right|^{2}\,.
\end{split}
\end{equation}
Here, $E^{\prime}=h\nu-E_{b}$ is the final energy obtained for a given optical frequency $\nu$ and binding energy $E_{b}$. Furthermore, $c$ is the speed of light, and $j'$ represents the rotational quantum number of the final level. We note that the transition moment $\langle \Lambda^{\prime};\,k|D_{\Lambda^{\prime},\Lambda}\left(R\right)|\Lambda\upsilon \rangle$ is essentially independent of $j$ and $j'$ for the low values of $j$ relevant here. From QCT calculations we estimate a typical range of rotational quantum numbers of $j<20$ for the BaRb$^+$ ion in our experiments.
$S\left(j^{\prime},j\right)$ denotes the H\"onl-London factor \cite{hansson2005}. In principle, transitions can be grouped into the three branches Q ($j^{\prime}=j$, $\Sigma$--$\Pi$ transitions only), R ($j^{\prime}=j+1$), and P ($j^{\prime}=j-1$). In our experiments we drive each of these transitions, if allowed by selection rules.  Summing over the P, Q, R contributions one obtains a total cross section which is independent of $j$. Therefore it is sufficient to present in the following  only total cross sections obtained for $j = 0$.
\begin{figure}[t]
\centering\includegraphics[width=\columnwidth]{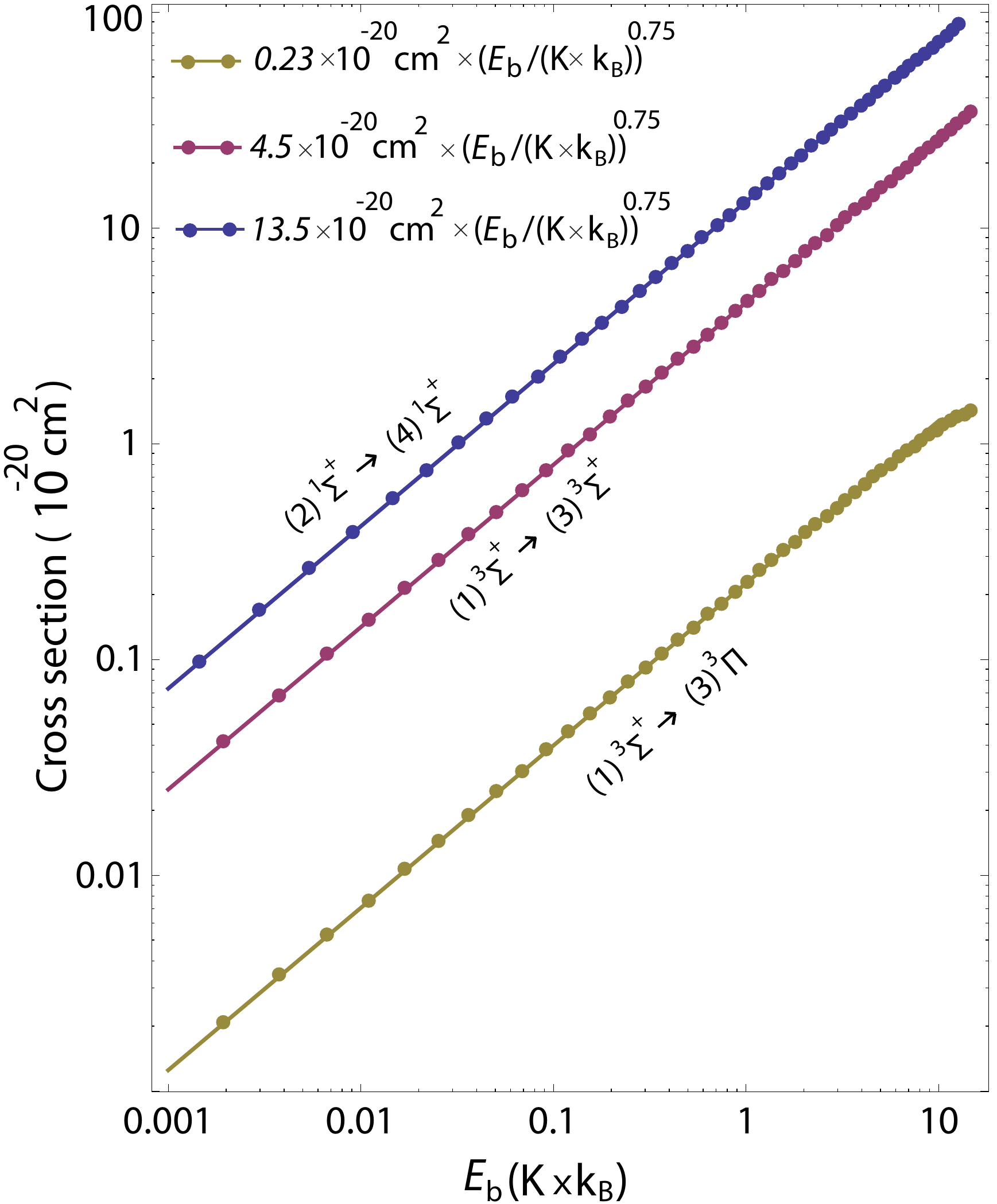}
\caption{Photodissociation cross sections as functions of the binding energy for an excited state BaRb$^+$ ion exposed to 1064$\:$nm light. Shown are the results for the most relevant transitions in our experiments. Data points are calculations. Solid lines represent fits $\propto E_\text{b}^{0.75}$ to the data points (see legend).}
\label{fig:CrossSectionVsBinEnergy}
\end{figure}

  Figure \ref{fig:CrossSectionVsBinEnergy} shows the predictions for photodissociation cross sections for 1064$\:$nm light as functions of the binding energy $E_b$ of the initial rovibrational state. Here, the three dominant transitions are presented. We checked numerically that the cross sections follow a $E_b^{0.75}$ scaling law within the shown range of $E_b$ \cite{Note8}. This can be explained by the increasing localization of the vibrational wave function with increasing binding energy $E_b$. Our calculations reveal that the transitions are mostly determined by the wave functions at the inner turning points of the PECs.

\begin{figure}[t]
\centering\includegraphics[width=\columnwidth]{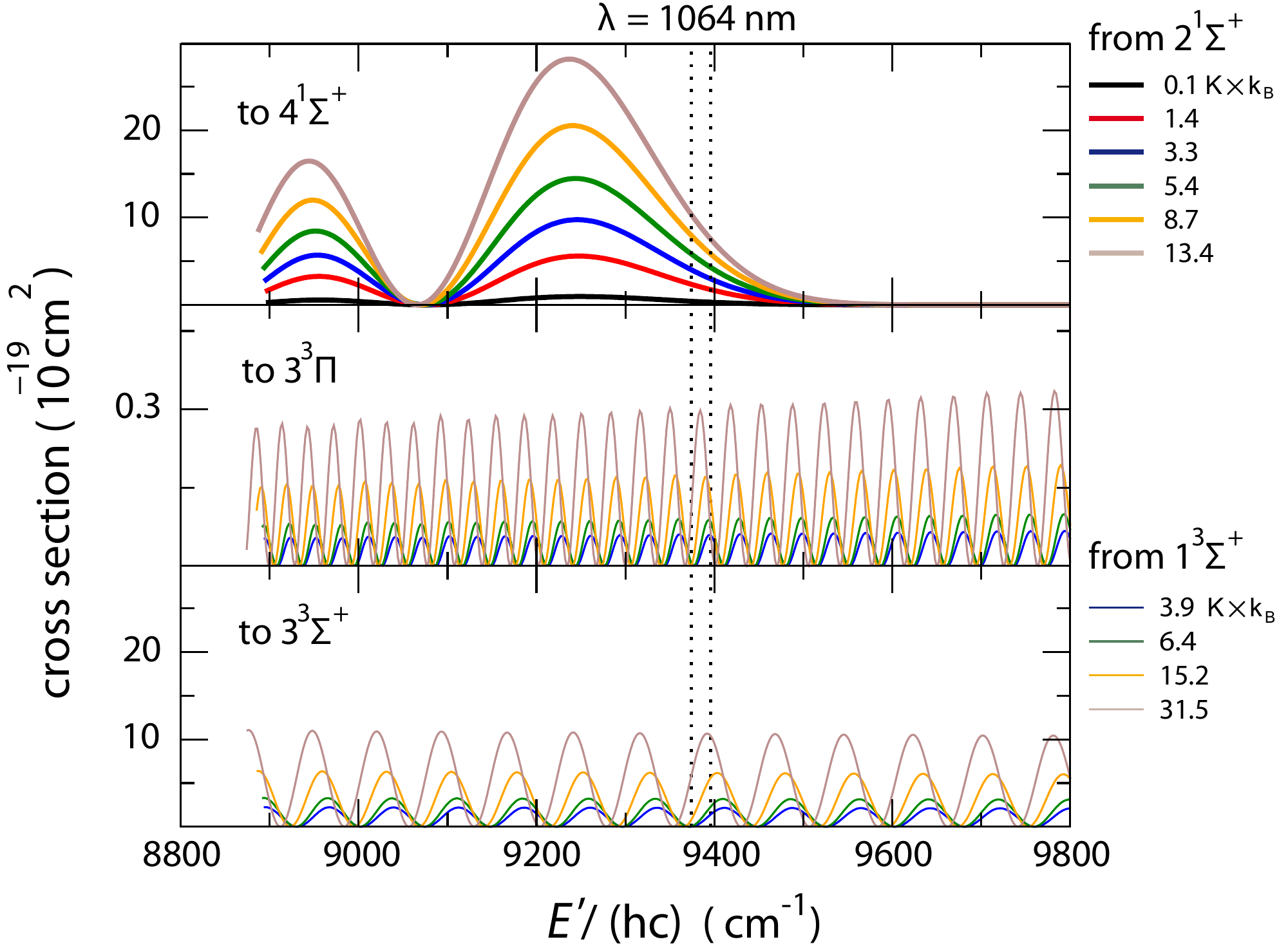}
\caption{Photodissociation cross sections as functions of the final energy $E^{\prime}=h\nu-E_{b}$. Upper panel: $\left(2\right)^{1}\Sigma^{+} \rightarrow (4)^{1}\Sigma^{+}$. Middle panel: $\left(1\right)^{3}\Sigma^{+}\rightarrow (3)^{3}\Pi$. Lower panel: $\left(1\right)^{3}\Sigma^{+}\rightarrow (3)^{3}\Sigma^+$. The color coding of the lines corresponds to the binding energy of the initial state as indicated on the right. The right (left) vertical dashed line marks the energy $E^\prime$ when a 1064$\:$nm photon excites a molecule with $E_b = 0.1\:\text{K}\times k_\text{B}$ ($E_b = 31.5\:\text{K}\times k_\text{B}$).
}
\label{fig:CrossSectionVsEnergy}
\end{figure}

We note that because of an uncertainty in the calculation of the absolute energy position of the PECs of up to a few hundred $\text{cm}^{-1} \times (hc)$ there are corresponding uncertainties in the photodissociation cross sections. The possible range of cross sections is investigated in Fig.$\:$\ref{fig:CrossSectionVsEnergy}. Here, $E^{\prime}/(hc)$ is varied between $8800$ and $9800\:\text{cm}^{-1}$, i.e. around typical values corresponding to final states addressed via light at 1064$\:$nm and starting from initial states with rather small binding energies (see dashed vertical lines). These calculations are carried out for several binding energies.

The cross sections in Fig.$\:$\ref{fig:CrossSectionVsEnergy} exhibit oscillations. For a fixed binding energy, the energy interval $\Delta E^\prime$ for a full oscillation between a minimum and a maximum is smaller than about $500\:\text{cm}^{-1}\times (hc)$ for all three presented transitions. This is about the uncertainty of the absolute energy positions of the PECs and therefore the true cross section can actually lie in the range between calculated minimum and maximum values.

\begin{figure}[t]
\centering\includegraphics[width=\columnwidth]{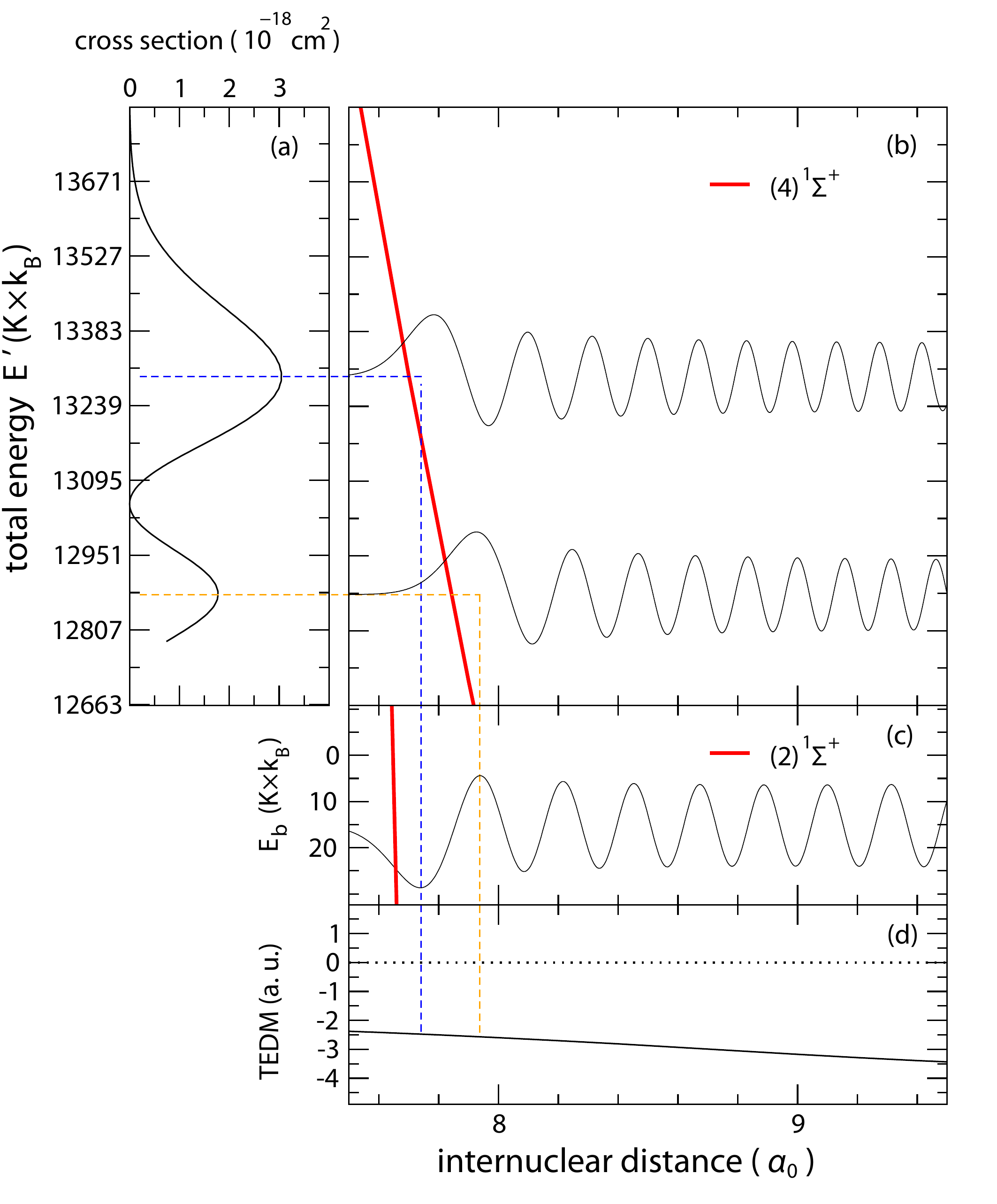}
\caption{(a) Photodissociation cross section as a function of final energy $E^{\prime}=h\nu-E_{b}$ for the transition $\left(2\right){}^{1}\Sigma^{+} \rightarrow \left(4\right){}^{1}\Sigma^{+}$. (b) The $\left(4\right){}^{1}\Sigma^{+}$ PEC (red curve) and two energy-normalized continuum wave functions (black curves). (c) The $\left(2\right){}^{1}\Sigma^{+}$ PEC (red curve) and the rovibrational wave function (black curve) for the binding energy of $E_{b}=10.5\:\text{cm}^{-1}$. Blue and orange dashed lines help to illustrate that a good wave function overlap at the inner turning point of the excited PEC leads to a large cross section.   (d) The respective TEDM as a function of the internuclear distance $R$.
}
\label{fig:PhotodissScheme}
\end{figure}

The oscillations of the cross sections in Fig.$\:$\ref{fig:CrossSectionVsEnergy} are associated with the spatial oscillations of the initial rovibrational wave functions. For the sake of clarity, this is illustrated in detail in Fig.$\:$\ref{fig:PhotodissScheme}. Wherever an anti-node of the initial wave function coincides with the anti-node of the scattering wave function at the inner turning point of the excited PEC, the cross section has a local maximum. This is known as the reflection principle (see, e.g., \cite{schinke1993}). The frequency separation of the local cross section maxima clearly depends on the slope of the PEC and the wavelength of the initial wave function.

In order to describe the experimentally measured data (see Appendix \ref{sec:fullevolution}), we use photodissociation cross sections
in our MC simulations of the form $\sigma_e \times (E_b / (\mbox{K}\times k_\text{B}))^{0.75}$. Thus, they exhibit the $E_b^{0.75}$ scaling, which our
calculations predict. The pre-factor $\sigma_e$, however, is used as free parameter which is
   determined via fits to the data.
    In Table \ref{tab:PDcsconstants} we compare
    the obtained values for $\sigma_e$  to the theoretically predicted maximal values. We find that the experimental cross section for the transition $(2\rightarrow 4)^1\Sigma^+$ (for the transition $(1)^3\Sigma^+ \rightarrow (3){^3\Pi}$) is by a factor of 13.5 (by a factor of 7) larger than the predicted maximal value. At this point it is not clear how to explain these discrepancies. In contrast, for the transition $(1 \rightarrow 3)^3\Sigma^+$ we find consistency between theory and experiment.

Having discussed in detail the photodissociation by $1064\,\text{nm}$ light, we now briefly comment on the photodissociation
by $493\,\text{nm}$ and $650\,\text{nm}$ light. Calculated PECs for highly excited electronic states (not shown here) indicate that the photodissociation of weakly-bound molecules in the $(2)^1\Sigma^+$ and $(1)^3\Sigma^+$ states
might be quite strongly suppressed because Condon points might not exist for the relevant transition. This agrees with the experiment, from which  we do not have any evidence for this photodissociation process either.
Concerning photodissociation of ground state $(X)^1\Sigma^+$ molecules via the laser cooling light, for which we do have experimental evidence (see Sec. \ref{sec:Photo_ground}), a theoretical analysis has not been carried out yet.

\begin{table}
\caption{Cross sections  $\sigma_e$ for the three transitions that are taken into account in our MC simulation. The predicted maximal values are given beside the values resulting from fits to the experimental data in our MC simulation. In the last column the released ion for each transition is given.}
\label{tab:PDcsconstants}
\begin{tabular}{l| c | c | c}
\hline
Transition  & max. $\sigma_e$ (theor.)  &    $\sigma_{e}$ (exp.)  & rel. ion \\
  &  ($10^{-20}\:$cm$^{2}$)  &   ($10^{-20}\:$cm$^{2}$)  & \\
  \hline
  $(2\rightarrow 4)^1\Sigma^+$ & 40  & 540 & Ba$^+$\\
  $(1 \rightarrow 3)^3\Sigma^+$ &  9  & 7 & Rb$^+$\\
  $(1)^3\Sigma^+ \rightarrow (3){^3\Pi}$ & 0.23 & 1.61 & Ba$^+$\\
  \hline
\end{tabular}
\end{table}

\subsubsection*{Radiative relaxation to the electronic ground state}
\label{sec:AppendRadRelax}

\begin{figure}[t]
\centering\includegraphics[width=0.7\columnwidth]{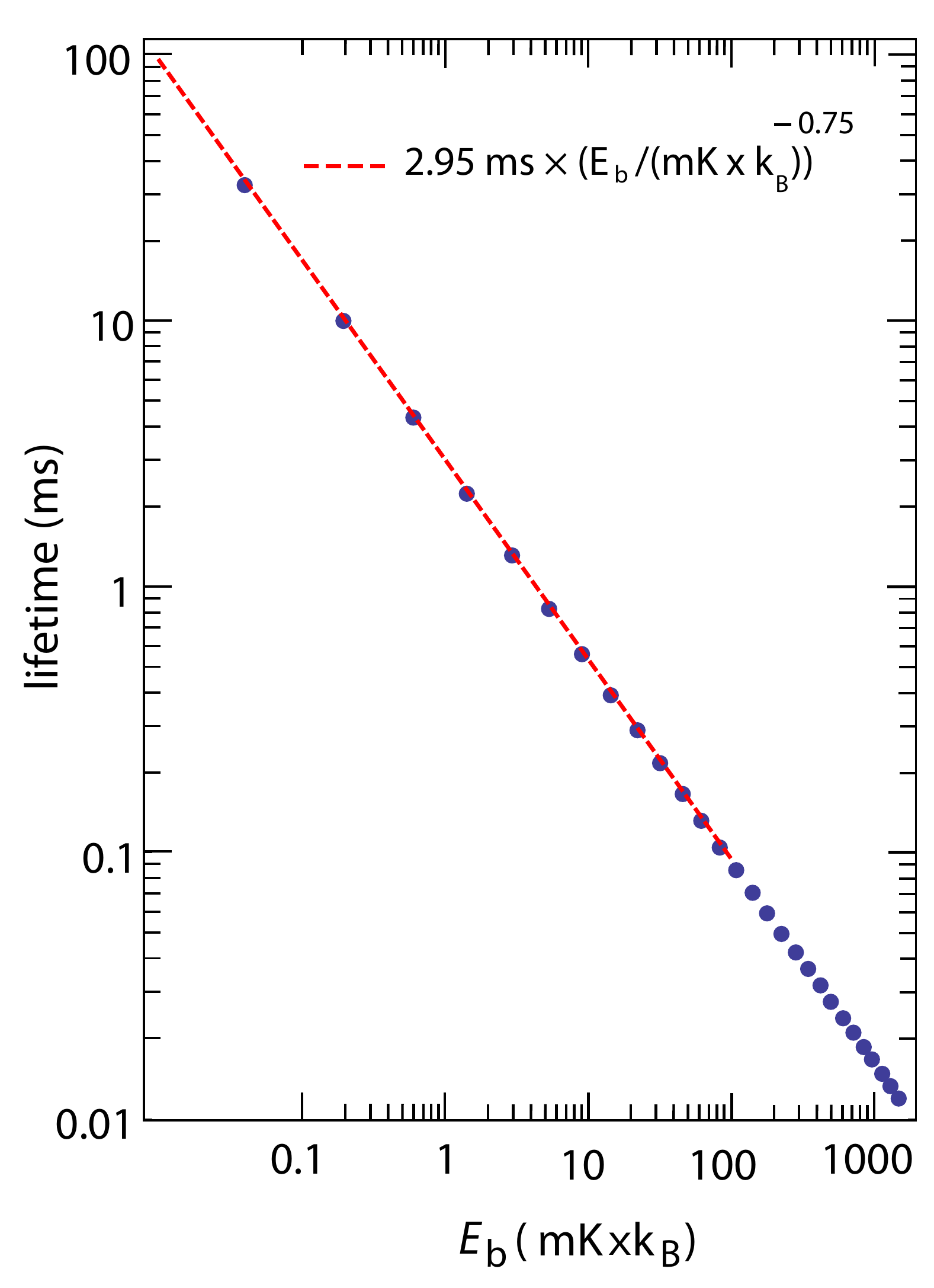}
\caption{Radiative lifetime (blue dots) for the highest rovibrational levels $\left(v,j=0\right)$ of the $\left(2\right)^{1}\Sigma^{+}$ electronic state, as a function of the binding energy $E_b$. The red dashed line is a fit $\propto E_b^{-0.75}$ to the data.}
\label{fig:LifetimeVsBinEnergy}
\end{figure}

\begin{figure}[h]
\centering\includegraphics[width=0.9\columnwidth]{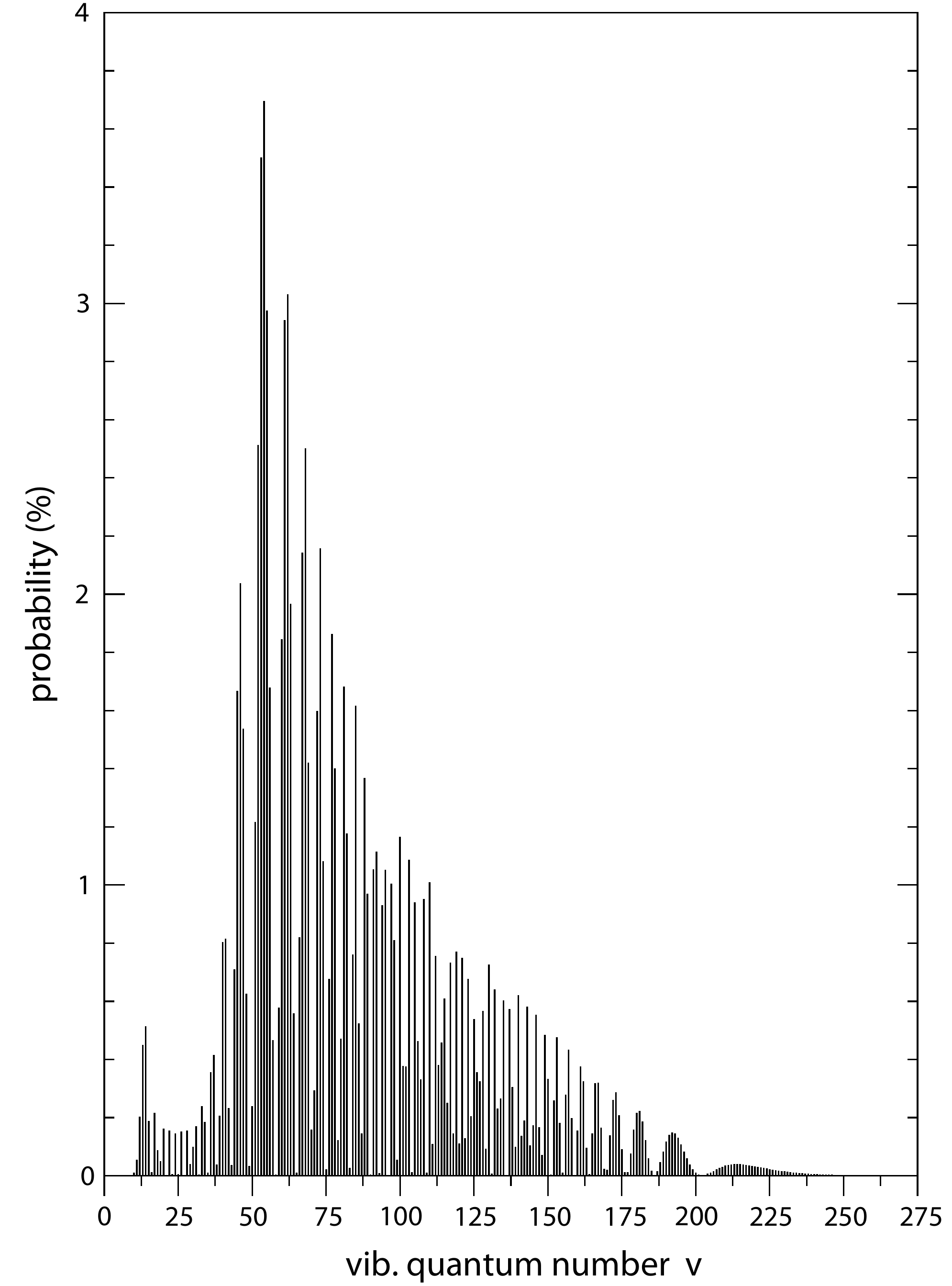}
\caption{Calculated population distribution for vibrational levels $v$ of the electronic ground state $(X)^1\Sigma^+$ of the BaRb$^+$ molecule after radiative relaxation from a weakly-bound level in the $(2)^1\Sigma^+$ state with $j=1$. Here, the same approach is used as described in \cite{Silva2015}.
}
\label{fig:vib_xbarb}
\end{figure}

The excited state $(2)^1\Sigma^+$ can  decay radiatively to the ground state $(X) ^{1}\Sigma^{+}$ by spontaneous emission of a photon.
 The corresponding radiative lifetime
 of the $(2)^{1}\Sigma^{+}$ molecule is shown in Fig.$\:$\ref{fig:LifetimeVsBinEnergy} as a function of the binding energy $E_b$, as previously discussed in \cite{Silva2015}. The relaxation can in principle lead to dissociation of the BaRb$^+$ molecule into a Rb$^+$ ion and a Ba atom. However, our calculations of the Franck-Condon factors show that it will dominantly produce a BaRb$^{+}$ molecule in the $(X)^{1}\Sigma^{+}$ state. Figure \ref{fig:vib_xbarb} shows the predicted broad distribution of vibrational levels which are populated.  During such a relaxation the kinetic  energy of the BaRb$^+$ molecule essentially does not change, because the photon recoil is very small. The radiative relaxation rate is the inverse of the lifetime, i.e. $ 0.34 \times (E_\text{b}/(\mbox{mK}\times k_\text{B}))^{0.75}\,\mbox{ms}^{-1}$ (see Fig.$\:$\ref{fig:LifetimeVsBinEnergy}). We use this relaxation rate in our MC simulations. The physics behind the scaling $\propto E_b^{0.75}$ is that for an increasing binding energy the wave function becomes more localized at short range where radiative relaxation dominantly occurs.

For the sake of completeness, we note that radiative relaxation within a given PEC (such as $(2)^1\Sigma^+$ or $(1)^3\Sigma^+$) is negligible in our experiments. As already discussed in \cite{Silva2015} these relaxation rates are on the order of seconds.

\section{ Monte Carlo Simulations}
\label{sec:App_MC}

In this section we describe how we simulate the evolution of a BaRb$^+$ molecule in a Rb atom cloud by means of MC trajectory calculations. For this, we make use of the cross sections we have determined in Appendix \ref{sec:app_ColCrossSec}. In order to reduce the complexity we carry out the calculations in two steps. In a first step we only consider a subset of collision processes. A main finding of these calculations is that the average kinetic energy of the BaRb$^+$  ion only slightly increases as it relaxes down to more deeply-bound vibrational states. We use this information in the second step of the MC calculations, where we now include all inelastic and reactive processes but for which we ignore elastic collisions and simply assume that the molecular ion has a constant kinetic energy.

\subsection{Restricted model }
\label{subs:Evolution}
Here, we simulate trajectories of a spinless BaRb$^+$ molecule. During each trajectory the molecule can undergo multiple collisions within the gas of Rb atoms. We consider elastic collisions, vibrational relaxation collisions, and collisional dissociation.

\begin{figure}[t]
	\centering\includegraphics[width=1.0\columnwidth]{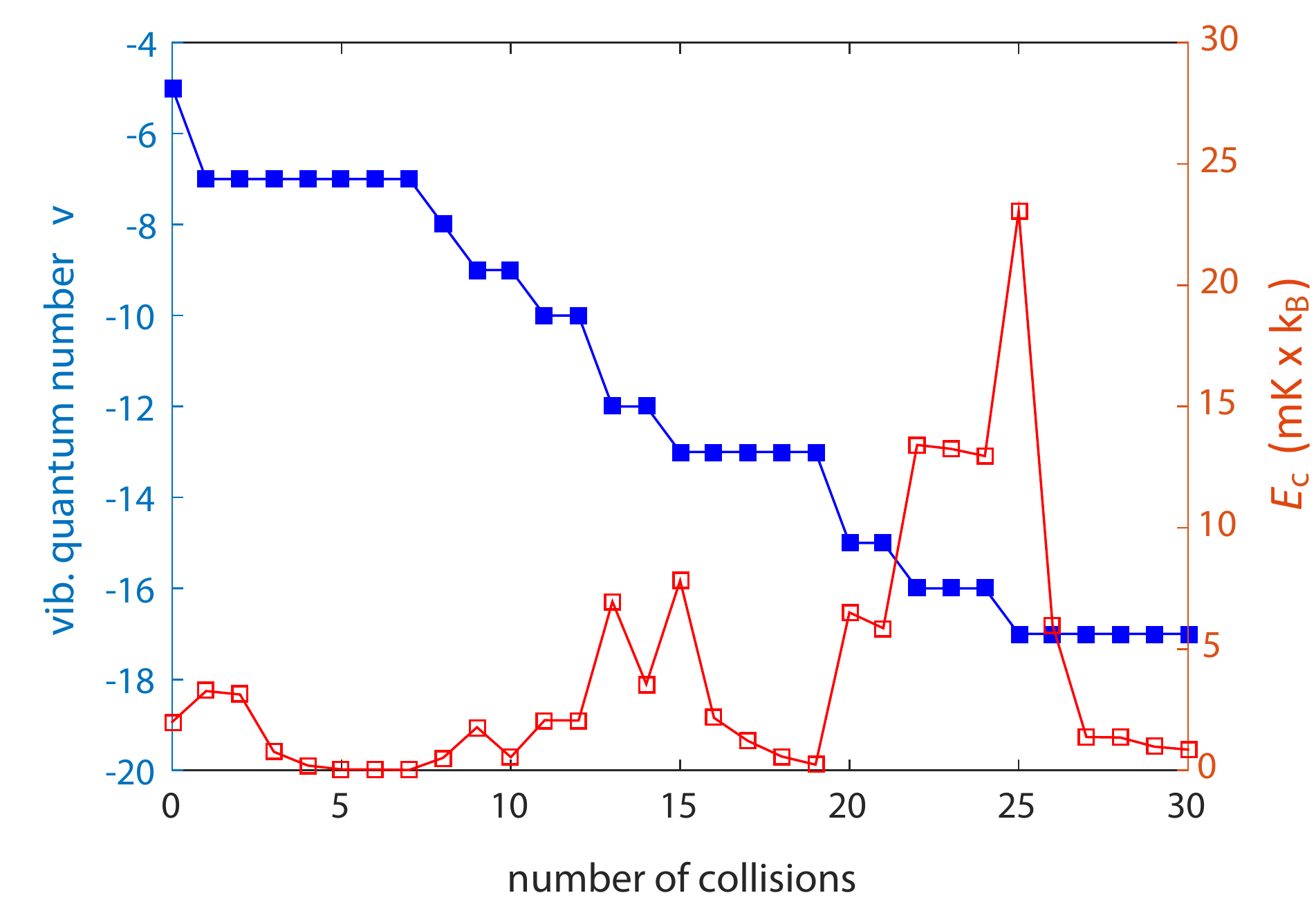}
	\caption{Evolution of the vibrational quantum number of a BaRb$^+$ molecule in collisions with Rb atoms. We only consider the "relevant" collisions which have impact parameters $b < b_{max}$, as discussed in Appendix \ref{sec:QTC}. The $y$-axis on the left shows the vibrational quantum number $v$ and the $y$-axis on the right shows the collision energy $E_\text{c}$. The data correspond to a single MC trajectory.}
	\label{fig:single_MC}
\end{figure}

\begin{figure}[t]
	\centering\includegraphics[width=0.85\columnwidth]{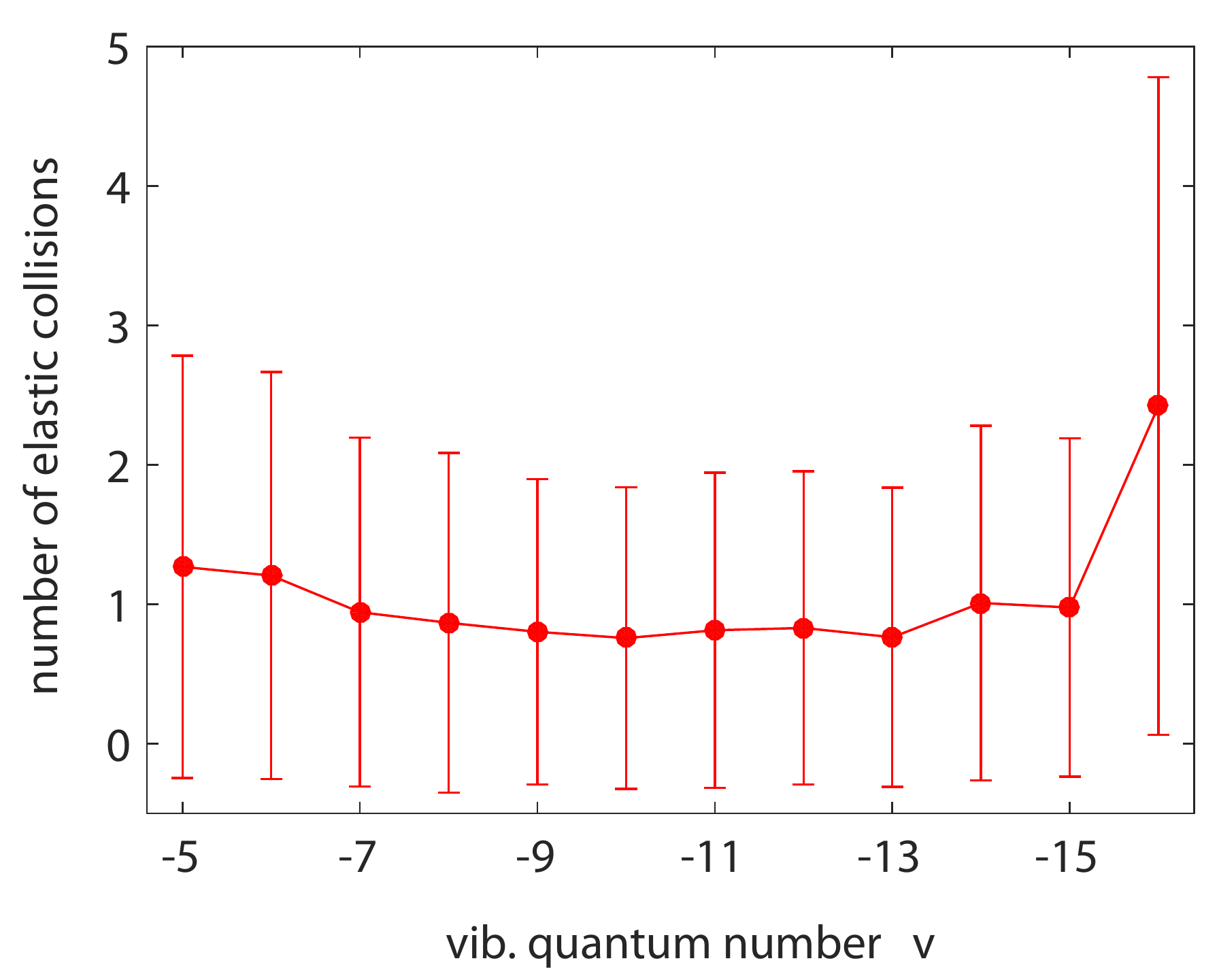}
	\caption{Average number of elastic collisions of Rb atoms with a BaRb$^+$ ion in the vibrational state $v$ before a relaxation (excitation) to another vibrational state occurs. Data points are from MC simulations. The error bars represent $1\sigma$ standard deviations.}
	\label{fig:elas_col}
\end{figure}

The simulation starts with the molecular ion in the vibrational state $v = -5$ below the $\text{Ba}^++\text{Rb}$ asymptote. An example of the evolution of the vibrational state as a function of the collision number for a single trajectory is shown in Fig.$\:$\ref{fig:single_MC}. The calculations reveal that vibrational relaxation typically takes place in steps of one or two vibrational quanta. The molecular kinetic energy increases after each vibrational relaxation step and decreases due to sympathetic cooling in elastic collisions. Precisely, how much energy is released in a vibrational relaxation step or carried away in an elastic collision depends on the scattering angle of the atom-molecule collision \cite{Landau1977}. In the simulations we choose random values for the scattering angle in the center-of-mass frame, which are uniformly distributed.

\begin{figure}[t]
	\centering\includegraphics[width=0.85\columnwidth]{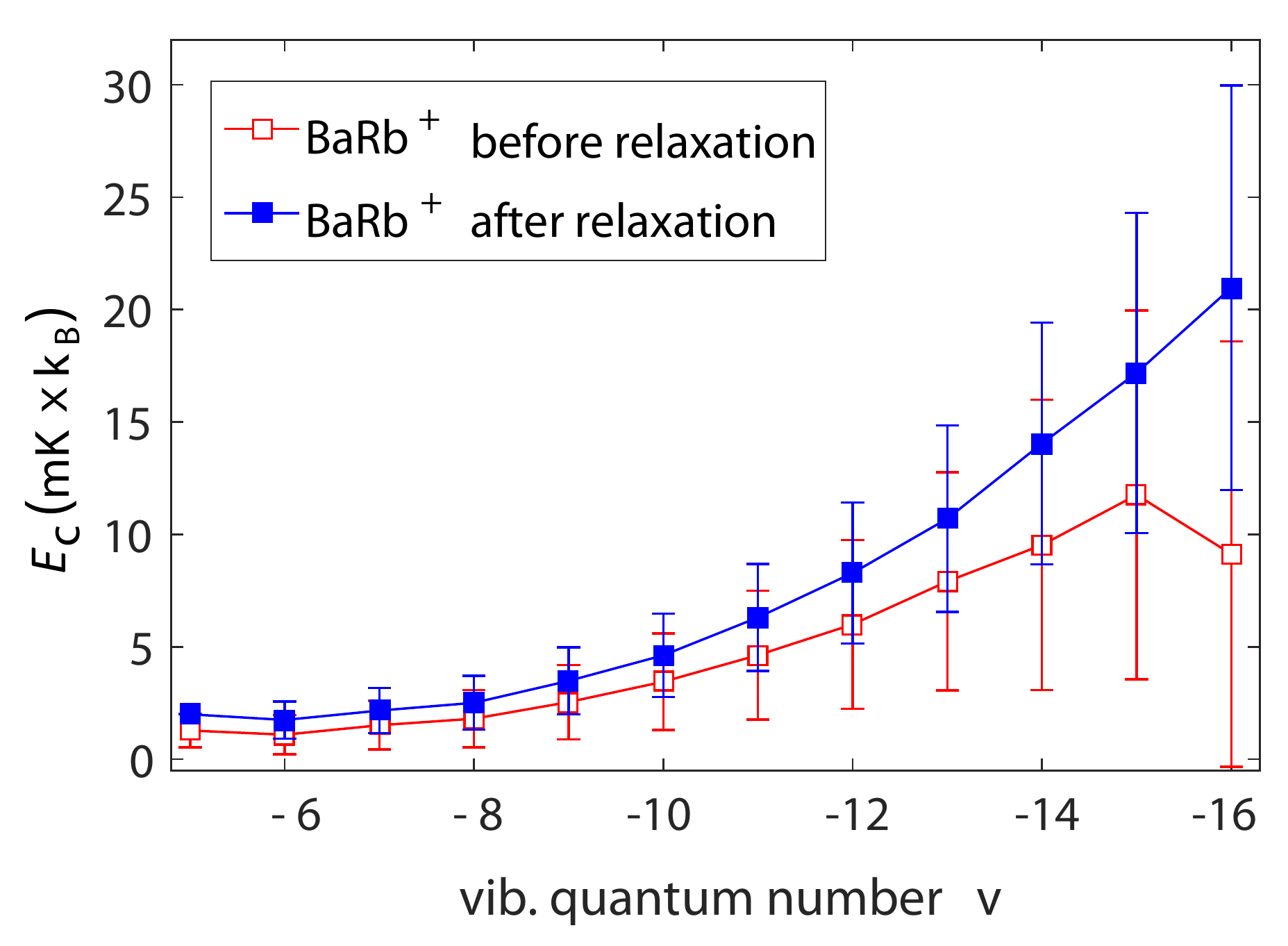}
	\caption{Average collision energy of the BaRb$^+$ ion right after (blue filled squares) and directly before (red open squares) vibrational relaxation as a function of the vibrational quantum number. Data points are results from the MC simulation. The error bars represent $1\sigma$ standard deviations. Here, the difference between the curves describes the effect of cooling due to elastic collisions.
	}	
	\label{fig:a_b_relax}
\end{figure}

\begin{figure}[t]
	\centering\includegraphics[width=0.9\columnwidth]{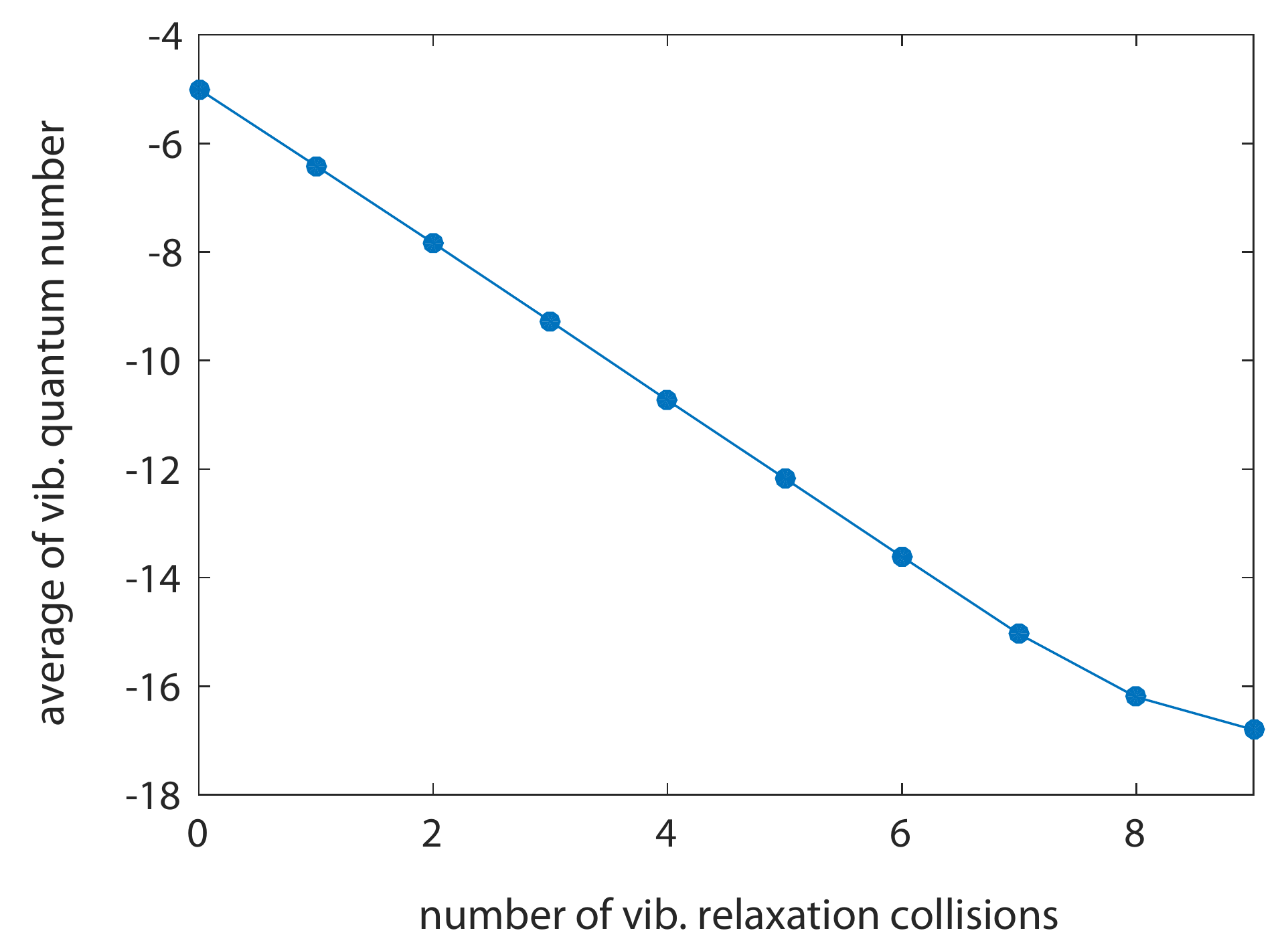}
	\caption{Average vibrational quantum number for a BaRb$^+$ ion as a function of the number of vibrational relaxation collisions with Rb atoms. The data points represent the results of MC calculations.	
    }
	\label{fig:Avg_vib_col}
\end{figure}

After analyzing 10$^4$ calculated trajectories we obtain the following results. Between two vibrational relaxation processes there is on average approximately one elastic collision (see Fig.$\:$\ref{fig:elas_col}). Although, overall, the kinetic energy of the molecule increases as it relaxes to more deeply-bound states, within the range of interest the molecular collision energies are typically only a few $\text{mK}\times k_\text{B}$ (see Fig.$\:$\ref{fig:a_b_relax}). On average, the molecular ion requires $17.5 \pm 4.2$ collisions to relax from $v=-5$ down to $v=-17$ of which $9.0 \pm 3.8$ collisions are elastic. Figure \ref{fig:Avg_vib_col} shows the average vibrational quantum number $v$ as a function of the number of vibrational relaxation collisions. We find that $v$ decreases nearly linearly.
 On average about 1.4 vibrational quanta are lost per relaxation collision,
  independent of the initial vibrational quantum number.
  Since the vibrational relaxation cross section is well approximated by the Langevin cross section (see Fig.$\:$\ref{fig:Vib_qunch}) the vibrational quantum number will on average be lowered by one unit at a rate of $1.4 \times \Gamma_\text{L}$. We note that also in a recent theoretical investigation of vibrational quenching collisions of weakly-bound Rb$_2^+$ molecular ions with Rb atoms the changes in the vibrational quantum number are predicted to be small \cite{Jachymski2020}.

\subsection{Full model}
\label{sec:fullevolution}
In the second set of MC trajectory calculations we take into account all the
processes discussed in Appendix \ref{sec:app_ColCrossSec}. Furthermore, we also include the formation of the weakly-bound BaRb$^+$ molecule with vibrational quantum number $v = -5$ via three-body recombination. Adopting simple statistical arguments and considering that the Ba$^+$ ion is initially unpolarized, the probability for the freshly formed molecule to be in state $(2)^1\Sigma^+$ ($(1)^3\Sigma^+$) is 1/4 (3/4), respectively.

\begin{figure}[t]
	\centering\includegraphics[width=\columnwidth]{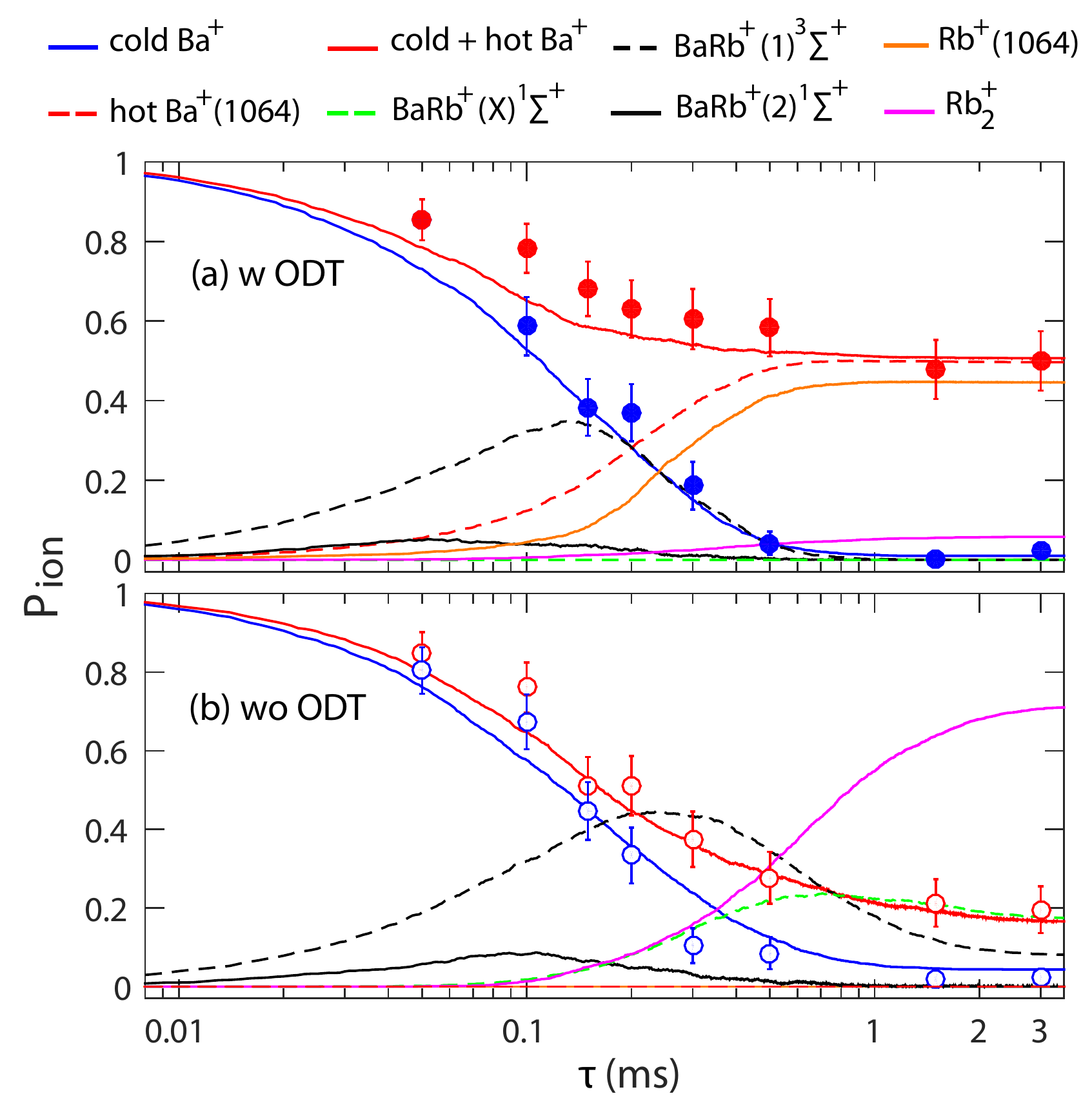}
	\caption{Calculations (lines) for ion signals together with measured data (filled and open circles) of Fig.$\:$\ref{fig2}(a). Shown are the probabilities for finding different ion states or species as given in the legend. In (a) the case with ODT is considered while in (b) the case without ODT is considered.
		"hot Ba$^+$(PD)" and "Rb$^+$(PD)" are populations due to photodissociation with 1064nm light. The population "Rb$_2^+$ / Rb$^+$" is due to secondary collisional reactions of BaRb$^+$ $(X)^1\Sigma^+$ molecules with Rb atoms.
	}
	\label{fig:all_MC_results}
\end{figure}

Motivated by the results in Appendix \ref{subs:Evolution} we generally assume a collision energy $E_c$ for the BaRb$^+$ molecule of a few mK$\times k_\text{B}$. For the collisional dissociation regarding the $v = -5$ level we assume $E_c=2\:\text{mK}\times k_\text{B}$. Actually, for all other collisional processes the precise value for the collision energy is not relevant since their rates are proportional to the Langevin rate, which is independent of $E_c$.
For the ground state $(X)^1\Sigma^+$, however, the assumption of low collision  energy is in general not justified.
This is because a vibrational relaxation from a deeply-bound vibrational level to the next one releases a large amount of energy. This puts the ion on an orbit through the Paul trap which is much larger than the size of the atom cloud. As a consequence the rate for further elastic, inelastic, or reactive collisions is significantly reduced.
For example, when the BaRb$^+$ molecule in state $(X)^1\Sigma^+$ relaxes from $v=55$ to $v=54$ the motional energy of the BaRb$^+$ molecule increases by about $16\,\text{K}\times k_\text{B}$. In order to get cooled back into the atom cloud, the energy has to be lowered to about $1\:\text{K} \times k_\text{B}$. On average, 44$\%$ of the energy is cooled away in a single elastic Langevin collision. Therefore, five elastic Langevin collisions are needed on average, to cool the BaRb$^+$ ion back down into the Rb gas. When we take this cooling time to be 130$\:\upmu \text{s}$ long we find good agreement with the data.

Once a highly energetic Ba$^+$, or a Rb$^+$, or a Rb$_2^+$ ion is produced after photodissociation or a substitution reaction, no further reaction takes place in our simulation. However, if a cold Ba$^+$ ion is created it can again undergo a three-body recombination event with the respective rate and a new evolution starts. Time is typically incremented in steps of $\Delta t=1\:\upmu\text{s}$. We typically carry out 2000 trajectories in a MC simulation for a given experimental procedure.

The results of the MC simulations are presented as lines in Figs.$\:$\ref{fig2}(a), \ref{fig3}, and \ref{fig:all_MC_results}. In fact, Fig.$\:$\ref{fig:all_MC_results} is an extension of Fig.$\:$\ref{fig2}(a), showing additional evolution traces for various ion states.  The measurements and predictions are shown separately for the case with dipole trap (wODT) and the case without dipole trap (woODT) in Fig.$\:$\ref{fig:all_MC_results}(a) and (b), respectively. The data points are the same as in Fig.$\:$\ref{fig2}(a).
 The plots clearly show how initially the populations of the $(2)^1\Sigma^+$ and $(1)^3\Sigma^+$ states increase due to formation of the BaRb$^+$ ion via three-body recombination. At some point later these populations peak and decrease due to radiative relaxation to the ground state $(X)^1\Sigma^+$ and, in the presence of $1064\:\text{nm}$ light, due to  photodissociation. The calculations for the creation of either a hot Ba$^+$ or a Rb$^+$ ion via this photodissociation are given by the curves denoted by hot Ba$^+$ (1064) and Rb$^+$ (1064), respectively. Radiative relaxation leads at first to a growing population of the $(X)^1\Sigma^+$ ground state BaRb$^+$ molecule which, in secondary reactions, is converted into a Rb$_2^+$ or a Rb$^+$ ion.  Here, we only consider the sum of the Rb$_2^+$ and Rb$^+$ populations, denoted Rb$_2^+$ / Rb$^+$.  When the $1064\:\text{nm}$ ODT is on, photodissociation is a dominant process for $(2)^1\Sigma^+$ and $(1)^3\Sigma^+$ molecules. Furthermore, the small fraction of molecules that relax to the ground state $(X)^1\Sigma^+$ are quickly removed in the trapped, dense atom cloud due to secondary reactions with Rb atoms.  In contrast, when the ODT is off, almost the whole ion population is first converted into ground state BaRb$^+$ molecules, apart from a small fraction remaining in the state $(1)^3\Sigma^+$. A sizable fraction of the ground state molecules do not undergo secondary reactions and therefore persist, as the released Rb atom cloud quickly falls away.
 In order to describe the experimental signal for "cold+hot Ba$^+$" we add the populations for "cold Ba$^+$" and "hot Ba$^+$(1064)" as well as 70\% of the population of BaRb$^+$ molecules in the states $(X)^1\Sigma^+$ and $(2)^1\Sigma^+$. This last contribution is due to photodissociation of ground state molecules. The scenario is the following. During the imaging all $(2) ^1\Sigma^+$ singlet molecules will relax to the ground state $(X)^1\Sigma^+$.  The cooling lasers will then dissociate these and the previously produced $(X)^1\Sigma^+ $ ground state  molecules. This photodissociation generates in 70\% (30\%) of the cases a hot Ba$^+$ (Rb$^+$) ion, as discussed in Sec.$\:$\ref{sec:Photo_ground}.


\begin{thebibliography}{28}
\bibitem [{Notev()}]{NoteVertLaser}%
We use the vertical ODT beam with an intensity of $1.8\:\text{kW}\,\text{cm}^{-2}$ which is only about 10\% of the intensity of the crossed ODT used for the experiments in Fig.$\:\ref{fig2}$. The light is switched on for 300$\:$ms which is about three orders of magnitude longer than the typical photodissociation time scale.
\bibitem [{Notev()}]{NoteAtomOffset}%
 About a quarter of the cold Ba$^+$ signal stems from experiments where due to a technical glitch no initial Rb cloud was produced and therefore the Ba$^+$ ion could never collide with Rb atoms.  These glitches occur in about 1\% of all runs.
\end{thebibliography}
\end{document}